\documentclass[letterpaper,conference,10pt]{IEEEtran}
\usepackage[margin=0.625in,right=0.650in,left=0.650in]{geometry}
\makeatletter
\def\ps@headings{%
\def\@oddhead{\mbox{}\scriptsize\rightmark \hfil \thepage}%
\def\@evenhead{\scriptsize\thepage \hfil \leftmark\mbox{}}%
\def\@oddfoot{}%
\def\@evenfoot{}}
\makeatother
\usepackage{subcaption}
\usepackage{amssymb,amsmath,amsthm, mathtools}
\usepackage{graphicx}
\usepackage{cite}
\usepackage{latexsym}
\usepackage{verbatim}
\usepackage{amsmath}
\usepackage{amssymb}
\usepackage{caption}
\usepackage{url}
\usepackage{color}
\usepackage{epstopdf}
\usepackage{algorithm}
\usepackage{algcompatible}
\usepackage[noend]{algpseudocode}
\usepackage{times}
\usepackage{stfloats}
\usepackage{multirow}
\usepackage{enumerate}
\usepackage{xspace}
\usepackage{longtable}
\usepackage{mathrsfs}
\usepackage{mathtools}
\usepackage{algcompatible}
\usepackage{fancyhdr}
\usepackage[noend]{algpseudocode}
\usepackage{amssymb}
\usepackage{pifont}
\usepackage{caption}
\usepackage{setspace}
\usepackage{bm}

\usepackage{enumitem}

\algrenewcommand\algorithmicindent{.5em}

\newcommand{\pir}{\ensuremath {\mathit{PIR}}{\xspace}}
\newcommand{\itpir}{\ensuremath {\mathit{itPIR}}{\xspace}}
\newcommand{\cpir}{\ensuremath {\mathit{cPIR}}{\xspace}}

\newcommand{\db}{\ensuremath {\mathit{DB}}{\xspace}} 

\newcommand{\crn}{\ensuremath {\mathit{CRN}}{\xspace}}
\newcommand{\sas}{\ensuremath {\mathit{SAS}}{\xspace}}

\newcommand{\su}{\ensuremath {\mathit{SU}}{\xspace}}
\newcommand{\pu}{\ensuremath {\mathit{PU}}{\xspace}}
\newcommand{\dbmatrix}{\ensuremath {\boldsymbol{\mathit{D}}}{\xspace}} 
\newcommand{\dbblock}{\ensuremath {\mathit{b}}{\xspace}} 
\newcommand{\dbrow}{\ensuremath {\mathit{r}}{\xspace}} 
\newcommand{\dbsize}{\ensuremath {\mathit{n}}{\xspace}} 
\newcommand{\dbshare}{\ensuremath {\mathit{\rho}}{\xspace}} 
\newcommand{\querymatrix}{\ensuremath {\boldsymbol{\mathit{Q}}}{\xspace}} 
\newcommand{\queries}{\ensuremath {\mathit{q}}{\xspace}} 

\newcommand{\ind}{\ensuremath {\mathit{\beta}}{\xspace}}
\mathchardef\mhyphen="2D 

\newcommand{\redundancy}{\ensuremath {\mathit{\pi}}{\xspace}}
\newcommand{\tp}{\ensuremath {\mathit{t}}{\xspace}} 
\newcommand{\kr}{\ensuremath {\mathit{k}}{\xspace}} 
\newcommand{\wnbr}{\ensuremath {\mathit{s}}{\xspace}} 
\newcommand{\ns}{\ensuremath {\mathit{\ell}}{\xspace}} 
\newcommand{\vbr}{\ensuremath {\mathit{\vartheta}}{\xspace}} 
\newcommand{\bitstr}{\ensuremath {\bm{\mathit{\rho}}}{\xspace}} 
\newcommand{\result}{\ensuremath {\bm{\mathit{R}}}{\xspace}} 
\newcommand{\word}{\ensuremath {\mathit{w}}{\xspace}} 
\newcommand{\x}{\ensuremath {\mathit{l_x}}{\xspace}} 
\newcommand{\y}{\ensuremath {\mathit{l_y}}{\xspace}} 
\newcommand{\chr}{\ensuremath {\mathit{C}}{\xspace}} 
\newcommand{\ts}{\ensuremath {\mathit{ts}}{\xspace}} 
\newcommand{\secret}{\ensuremath {\mathcal{S}}{\xspace}} 
\newcommand{\scode}{\ensuremath {\mathit{S}}{\xspace}} 

\newcommand{\PrSpec}{\ensuremath {\mathit{PriSpectrum}}{\xspace}}
\newcommand{\chorScheme}{\ensuremath {\mathit{LP\mhyphen Chor}}{\xspace}}
\newcommand{\goldbergScheme}{\ensuremath {\mathit{LP\mhyphen Goldberg}}{\xspace}}
\newcommand{\batchScheme}{\ensuremath {\mathit{LP\mhyphen BatchPIR}}{\xspace}}

\newcommand{\algrule}[1][.2pt]{\par\vskip.5\baselineskip\hrule height #1\par\vskip.5\baselineskip}

\newtheorem{mycorollary}{Corollary}{\bfseries}{\rmfamily}

{\bfseries}{\rmfamily}
\newtheorem{definition}{Definition}{\bfseries}{\rmfamily}
{\bfseries}{\rmfamily}

\IEEEoverridecommandlockouts
\begin{document}

\title{
Location Privacy in Cognitive Radios with Multi-Server Private Information Retrieval }


\author{Mohamed Grissa, Attila A. Yavuz, and Bechir Hamdaoui\\
\small Oregon State University, grissam,hamdaoui@oregonstate.edu\\
\small University of South Florida, attilaayavuz@usf.edu
}

\maketitle

\begin{abstract}
Spectrum database-based cognitive radio networks (\crn s) have become the de facto approach for enabling unlicensed secondary users (\su s) to identify spectrum vacancies in channels owned by licensed primary users (\pu s). Despite its merits, the use of spectrum databases incurs privacy concerns for both \su s and \pu s. Single-server private information retrieval (\pir) has been used as the main tool to address this problem. However, such techniques incur extremely large communication and computation overheads while offering only computational privacy. Besides, some of these \pir~ protocols have been broken.

In this paper, we show that it is possible to achieve high efficiency and (information-theoretic) privacy for both \pu s and \su s in database-driven \crn~with multi-server \pir. Our key observation is that, by design, database-driven \crn s comprise multiple databases that are required, by the Federal Communications Commission, to synchronize their records. To the best of our knowledge, we are the first to exploit this observation to harness multi-server \pir~technology to guarantee an optimal privacy for both \su s and \pu s, thanks to the unique properties of database-driven \crn. We showed, analytically and empirically with deployments on actual cloud systems, that multi-server \pir~is an ideal tool to provide efficient location privacy in database-driven \crn.
\end{abstract}


\section{Introduction}
\label{sec:Introduction}
The rapid growth of connected wireless devices has dramatically increased the demand for wireless spectrum and led to a serious shortage in spectrum resources. Cognitive radio networks (\crn s)~\cite{mitola1999cognitive} have emerged as a promising technology for solving this shortage problem by enabling dynamic spectrum access (DSA), which improves the spectrum utilization efficiency by allowing unlicensed/secondary users (\su s) to exploit unused spectrum bands (aka spectrum holes or white spaces) of licensed/primary users (\pu s).

Currently, two approaches are being adopted to identify these white spaces: spectrum sensing and geolocation spectrum databases. In the spectrum sensing-based approach, \su s need to sense the \pu~channel to determine whether the channel is available for opportunistic use. The spectrum database-based approach, on the other hand, waives the sensing requirement and instead enables \su s to query a database (\db) to learn about spectrum opportunities in their vicinity. This approach, already promoted and adopted by the Federal Communications Commission (FCC), was introduced as a way to overcome the technical hurdles faced by the spectrum sensing-based approaches, thereby enhancing the efficiency of spectrum utilization, improving the accuracy of available spectrum identification, and reducing the complexity of terminal devices~\cite{gao2013location}. Moreover, it pushes the responsibility and complexity of complying with spectrum policies to \db~and eases the adoption of policy changes by limiting updates to just a handful number of databases, as opposed to updating large numbers of devices~\cite{chen2015protocol}.

FCC has designated nine entities (e.g. Google~\cite{google}, iconectiv~\cite{iconectiv}, and Microsoft~\cite{microsoft}) as TV bands device database administrators which are required to follow the guidelines provided by PAWS (Protocol to Access White Space) standard~\cite{chen2015protocol}. PAWS sets guidelines and operational requirements for both the spectrum database and the \su s querying it. These include: \su s need to be equipped with geo-location capabilities, \su s must query \db~with their specific location to check channel availability before starting their transmissions, \db~must register \su s and
manage their access to the spectrum, \db~must respond to \su s' queries with the list of available channels in their vicinity along with the appropriate transmission parameters. As specified by PAWS standard, \su s may be served by several spectrum databases and are required to register to one or more of these databases prior to querying them for spectrum availability. The spectrum databases are reachable via the Internet, and \su s querying these databases are expected to have some form of Internet connectivity\cite{mancuso2013protocol}.

FCC has established a new service in the 3.5 GHz band, known as Citizens Broadband Radio Service (CBRS), in which the spectrum is also managed through a central database-driven \crn, aka spectrum access system (\textsl{SAS}), to enable spectrum sharing between military and federal incumbents and \su s. A separate entity with Environmental Sensing Capability (ESC) is responsible of populating \db s with data regarding \pu s that do not wish to reveal their operational information such as their location or transmission characteristics. A similar concept, named licensed shared access (LSA), for the 2.3-3.4 GHz band is also being developed in Europe to enable \su s to opportunistically access spectrum resources in this band owned by incumbent military aircraft services and police wireless communications. A major difference compared to SAS, is that in LSA, \pu s are responsible for populating \db s by providing their a priori information; i.e. their activities and, therefore the spectrum availability information, are known upfront~\cite{massaro2017next}.

%

\subsection{Location Privacy Issues in Database-Driven \crn s}

Despite their benefits, database-driven \crn s suffer from serious security and privacy threats. Since they could be seen as a variant of of {\em location based service (LBS)}, the disclosure of location information of \su s represents the main threat to \su s when it comes to obtaining spectrum availability from \db s. The fine-grained location, when combined with publicly available information, can easily reveal other personal information about an individual including his/her behavior, health condition, personal habits or even beliefs. For instance, an adversary can learn some information about the health condition of a user by observing that the user regularly goes to a hospital for example. The frequency and duration of these visits can even reveal the seriousness of a user illness and even the type of illness if the location corresponds to that of a specialty clinic. Matters get worse when \su s are mobile. As per the PAWS requirements, \su s need to query \db s whenever they change their location by at least 100 meters. This will make \su s constantly share their location as they move which could be exploited by a malicious service provider for tracking purposes. 

The location privacy of \su s is not the only privacy concern that database-driven \crn s suffer from. Indeed, the location privacy of \pu s may also be critical in \crn~systems such as \sas, in the 3.5 GHz CBRS band, and LSA, in the 2.3-2.4 GHz band, where \pu s are not commercial but rather military and governmental entities. To achieve efficient spectrum sharing without interference to military and federal incumbents, these systems~require \pu s, or entities with sensing capabilities such as ESC, to report \pu s' operational data (including their location, frequencies time of use, etc.) to be included in the spectrum databases which may present serious privacy risks to these \pu s.


Being aware of such potential privacy threats, both \su s and \pu s may refuse to share their sensitive information with \db s, which may present a serious barrier to the adoption of database-based \crn s, and to the public acceptance and promotion of the dynamic spectrum sharing paradigm. Therefore, {\em there is a critical need for developing techniques to protect the location privacy of both \pu s and \su s while allowing the latter to harness the benefits of the \crn~paradigm without disrupting the functionalities that these techniques are designed for to promote dynamic spectrum sharing}.


\subsection{Research Gap and Objectives} \label{subsec:ResearchGapObjectives}

Despite the importance of the location privacy issue in \crn s, only recently has it started to gain interest from the research community~\cite{grissa2017location}. Some works focus on addressing this issue in the context of collaborative spectrum sensing~\cite{li2012location,grissa2015location,wangprivacy,grissa2016efficient,grissa2017preserving}; others address it in the context of dynamic spectrum auction~\cite{liu2013location}. 
%
Protecting \su s' location privacy in database-driven \crn s is a more challenging task, merely because \su s are required, by protocol design, to provide their physical location to \db~to learn about spectrum opportunities in their vicinity. The heterogeneity of wireless devices and the versatility of services relying on the CRN technology~\cite{wang2016software} could also present some challenges in designing privacy-preserving mechanisms for users in \crn s. In fact, privacy-preserving solutions need to embrace the different resource constraints of each \su~device and the various requirements of each service in terms of data rates and delay sensitivities. This makes it hard to leverage general purpose public key encryption-based techniques due to their high cost in terms of computation and communication overheads especially on resource-constrained devices. It is therefore crucial to design cost-effective protocols that offer strong privacy guarantees to users and also adapt to different systems requirements regardless of the constraints of the users.

The existing location privacy preservation techniques for database-driven \crn~(e.g.,~\cite{zhang2015optimal,gao2013location,troja2014leveraging,troja2015efficient,zhang2015achieving,grissa2017locationPrivacy}) generally rely on three main lines of privacy preserving technologies, (i)   {\em k-anonymity}~\cite{gruteser2003anonymous}, (ii) {\em differential privacy}~\cite{dwork2008differential}  and (iii) single-server {\em Private Information Retrieval (\pir)}~\cite{chor1998private}. However, the direct adaptation of {\em k-anonymity} based techniques have been shown to yield either insecure or extremely costly results~\cite{zang2011anonymization}. The solutions adapting {\em differential privacy} (e.g.,~\cite{zhang2015achieving}) not only incur a non-negligible overhead, but also introduce a noise over the queries, and therefore they may negatively impact the accuracy of spectrum availability information.

Among these alternatives, single-server \pir~seems to be the most popular. \pir~technology is a suitable choice for database-driven \crn s, as it permits privacy preserving queries on a public database, and therefore can enable a \su~to retrieve spectrum availability information from the database without leaking its location information. However, single-server \pir~protocols rely on highly costly partial homomorphic encryption schemes, which need to be executed over the entire database for each query. Indeed, as we also demonstrated with our experiments in Section \ref{sec:Performance}, the execution of a single query even with some of the most efficient single-server \pir~schemes~\cite{aguilar2016xpir} takes approximately $20$ seconds with a $80\:Mbps/\:30Mbps$ bandwidth on a moderate size database (e.g., $10^6$ entries). An end-to-end delay with the orders of $20$ seconds might be undesirable for spectrum sensing needs of \su s in real-life applications. Also, some of the state-of-the-art efficient computational \pir~schemes~\cite{trostle2010efficient} that are used in the context of \crn s have been shown to be broken~\cite{aguilar2016xpir}. Thus, there is a significant need for practical location privacy preservation approaches for database-driven \crn s that can meet the efficiency and functionality requirements of \su s.

\subsection{Our Observation and Contribution}
The objective of this paper is to develop efficient techniques for database-driven \crn s that preserve the location privacy of \su s during their process of acquiring spectrum availability information. We also try to protect the operational privacy of \pu s in systems that require incumbents to provide spectrum availability information to \db s. Specifically, we will aim for the following design objectives: $(i)$ ({\em location privacy of \su s}) Preserve the location privacy of \su s, whether fixed or mobile, while allowing them to receive spectrum availability information; $(ii)$ {\em (efficiency and practicality)} Incur minimum computation, communication and storage overhead. The cryptographic delay must be minimum to permit fast spectrum availability decision for the \su s, and storage/processing cost must be low to enable practical deployments. $(iii)$ {\em (fault-tolerance and robustness)} Mitigate the effects of system failures or misbehaving entities (e.g., colluding databases). $(iv)$ {\em (location privacy of \pu s) } The location information of \pu s needs to be protected while still able to provide spectrum availability information to \db s. {\em It is very challenging to meet all of these seemingly conflicting design goals simultaneously.}

The main idea behind our proposed approaches is to harness special properties and characteristics of the database-driven \crn~systems to employ private query techniques that can overcome the significant performance, robustness and privacy limitations of the state-of-the-art techniques. Specifically, our proposed approach is based on the following observation:
%

\textbf{Observation}: {\em FCC requires that all of its certified databases synchronize their records obtained through registration procedures with one another~\cite{fcc2012white,fcc2012TVWS} and need to be consistent across the other databases by providing exactly the same spectrum availability information, in any region, in response to \su s' queries~\cite{ramjee2016critique}. That is, the same copy of spectrum database is available and accessible to the \su s via multiple (distinct) spectrum database administrators/providers.  Is it possible exploit this observation to achieve efficiency location preservation techniques for database-driven \crn?}

In practice,  as stated in PAWS standard~\cite{chen2015protocol}, \su s have the option to register to multiple spectrum databases belonging to multiple service providers.  Currently, many companies (e.g. Google~\cite{google}, iconectiv~\cite{iconectiv}, etc) have obtained authorization from FCC to operate geo-location spectrum databases upon successfully complying to regulatory requirements. Several other companies are still underway to acquire this authorization\cite{fcc-administrators}. Thus, it is more natural and realistic to take this fact into consideration when designing privacy preserving protocols for database-based \crn s. Based on this observation, our main contribution is as follows:
\begin{table}[h!]
\centering  \caption{ Performance Comparison} \label{tab:Table0}

\renewcommand{\arraystretch}{1.5}{
\resizebox{.5\textwidth}{!}{%
\begin{tabular}{||c||c||c|c|c||c||}

\hline {\multirow{2}{*}{\textbf{Scheme} }}   & {\multirow{2}{*}{\textbf{Comm.}}} & \multicolumn{3}{|c||}{\textbf{Delay}} & {\multirow{2}{*}{\textbf{Privacy}}}\\ \cline{3-5}
 &  &  {$\boldsymbol \db$} & {$\boldsymbol \su$}& \textbf{ total}& \\ \hline

\hline \hline  \chorScheme &  $ 753\:KB $ &  $0.48\:s$ & $0.0077\:s$ & $0.62\:s$& $(\ns-1)${\em-private}\\

\hline  \goldbergScheme & $ 6000\:KB$ & $1.21\:s $ & $ 0.32\: s$ & $1.78\:s$& \tp{\em-private \ns-comp.-private} \\

\hline  $RAID\mhyphen$\chorScheme & $125\:KB$ & $0.022\:s$ & $ 0.00041\:s$ & $0.21\:s$& $(\redundancy-1)${\em-private} \\

\hline  \hline \PrSpec~\cite{gao2013location} & $ 512.8\: KB $  & $21\:s $ & $0.084 \:s$& $24.2$ & {\em underlying \pir~ broken}\\

\hline  Troja et al~\cite{troja2015efficient} & $ 8.4\:KB$ & $11760\:s$ & $5.62\:s$ & $11766\:s$ & {\em computationally-private}\\

\hline  Troja et al~\cite{troja2014leveraging} & $12120\:KB $ & $11760\:s$ & $48\:s$& $11820\:s$ & {\em computationally-private}\\

\hline \hline XPIR~\cite{aguilar2016xpir} & $ 4321\:KB $ & $17.66\:s$ & $ 0.34\:s$ & $20.53\:s$ & {\em computationally-private}\\ \hline

\hline \hline SealPIR~\cite{angel2018pir} & $ 512\:KB$ & $ 11.03\:s$ & $0.008\:s $ & $ 11.35\:s $ &\em computationally-private\\ \hline 
\end{tabular}}
}
\begin{center}
  \scriptsize{\textbf{Parameters:} $\dbsize = 560\:MB,\:\dbblock = 560\:B, \:\dbrow = 10^6, \:\ns=6, \:\word = 8, \;\kr = 6$
\vspace{-6pt}
}  
\end{center}

\end{table}


\textbf{Our Contribution}: {\em To the best of our knowledge, we are the first to exploit the fact that multiple copies of spectrum \db s are available by nature in database-driven \crn s, and therefore it is possible to harness  multi-server \pir~techniques~\cite{chor1998private,goldberg2007improving} that offer information-theoretic privacy with substantial efficiency advantages over single-server \pir. This is achieved by relying on Shamir secret sharing-based techniques to either divide the content of \su s' queries or the spectrum availability information, or both, among the different \db s to prevent these \db s from inferring \su s' location from their queries or from learning \pu s' sensitive operational data from the spectrum availability information.

We show, analytically and experimentally with deployments on cloud systems, that our adaptation of multi-server \pir~techniques significantly outperforms the state-of-the-art location privacy preservation methods as demonstrated in Table~\ref{tab:Table0} and detailed in Section \ref{sec:Performance}. Moreover, our adaptations achieve information theoretical privacy while existing alternatives offer only computational privacy. This feature provides an assurance against even post-quantum adversaries~\cite{chen2016nistir} and can avoid recent attacks on computational \pir~\cite{aguilar2016xpir}.}

Notice that, multi-server \pir~techniques require the availability of multiple (synchronized) replicas of the database. Therefore, despite their high efficiency and security, they received a little attention from the practitioners. For instance, in traditional data outsourcing settings (e.g., private cloud storage), the application requires a client to outsource only a single copy of its database. The distribution and maintenance of multiple copies of the database across different service providers brings additional architectural and deployment costs, which might not be economically attractive for the client.

In this paper, we showcased one of the first natural use-cases of multi-server \pir, in which the multiple copies of synchronized databases are already available by the original design of application (i.e., spectrum availability information in multi-database \crn s), and therefore multi-server \pir~does not introduce any extra overhead on top of the application. Exploiting this synergy between multi-database \crn~and multi-server \pir~permitted us to provide informational theoretical location privacy for \su s with a significantly better efficiency compared to existing single-server \pir~approaches.

\textbf{Desirable Properties:} We outline the desirable properties of our approaches below.

\begin{itemize}[leftmargin=*]
\item \noindent{\em Computational efficiency:} The adapted approaches are much more efficient than existing location privacy preserving schemes. For instance, as shown in Table~\ref{tab:Table0}, \chorScheme~and \goldbergScheme~are more than $3$ orders of magnitudes faster than the schemes proposed by Troja et al.~\cite{troja2014leveraging,troja2015efficient}, and $10$ times faster than XPIR~\cite{aguilar2016xpir} and \PrSpec~\cite{gao2013location}.
\item {\em Information Theoretical Privacy Guarantees:} They can achieve information-theoretic privacy which is the optimal privacy level that could be reached as opposed to computational privacy guarantees offered by existing approaches. In fact some of these approaches are prone to recent attacks on computational-\pir~protocols~\cite{aguilar2016xpir} and are not secure against post-quantum adversaries~\cite{chen2016nistir}.
\item {\em Low communication overhead:} Our approaches incur a reasonable communication overhead that is a middle ground between the fastest computational \pir~\cite{aguilar2016xpir} and the most communication efficient computational \pir~\cite{gentry2005single}.
    \item {\em Fault-Tolerance and Robustness}: Our proposed approaches are resilient to the issues that are associated with multi-server architectures: failures, byzantine behavior, and collusion. Even though the collusion of all of the service providers is unlikely to happen due to the competing nature of these companies and due to regulatory enforcement from bodies such as FCC to protect users’ data, we have however considered collusion in our system and security model.
    All proposed approaches can handle collusion of multiple \db s up to certain limit that is different for each approach. In addition, some of the proposed approaches can also handle faulty and byzantine~\db s. Besides, simply hacking \db s, when the proposed approaches are in place, will not be sufficient to learn users' information since some of these protocols offer hybrid privacy protection by combining both computational~and information-theoretic \pir~protocols enabling them to offer computational privacy even when all of the \db s are compromised.
 \item {\em Experimental evaluation on actual cloud platforms}: We deploy our proposed approaches on a real cloud platform, GENI~\cite{Berman20145}, to show their feasibility. In our experiment, we create multiple geographically distributed VMs each playing the role of a \db. A laptop plays the role of a \su~that queries \db s, i.e. VM s. Our experiments confirm the superior computational advantages of the adoption of multi-server \pir~over the existing alternatives.
\end{itemize}

\subsection{Differences Compared to the Preliminary Version}
The main differences between this paper and its preliminary versions~\cite{grissa2017when,grissa2018unleashing} are as follows: (i) We further consider the location privacy issue of mobile \su s and offer a way to amortize the cost incurred by mobility. (ii) We also leverage multi-server \pir~to address the location privacy issue of \pu s in database-\crn~systems that require \pu s to provide spectrum availability to \db s. (iii) We discuss also a way to reduce the cost of \chorScheme~by partitioning the spectrum database instead of simply replicating it using the RAID-PIR protocol~\cite{demmler2014raid} and we discuss the privacy-performance tradeoff of relying on such approach. (iv) We provide a more detailed performance evaluation that takes into account the latest advances in \pir~technology, namely SealPIR~\cite{angel2018pir} which relies on fully homomorphic encryption.

\section{Preliminaries and Models}
\label{sec:Prelim}
\subsection{Notation and Building Blocks}
\label{sec:blocks}
We summarize our notations in Table~\ref{t:notations}. Our adaptations of multi-server \pir~rely on the following building blocks.

\begin{table}[h!]
\caption{\small Notations}
\centering
\resizebox{0.4\textwidth}{!}{
\label{t:notations}
\begin{tabular}{l l}

\hline
\noalign{\medskip }
$\db$ & Spectrum database \\
$\su$ & Secondary user \\
$\crn$ & Cognitive radio network \\
$\ns$ & Number of spectrum databases \\
$\dbmatrix$ & Matrix modeling the content of \db \\
$\dbrow$ & Number of records in \dbmatrix \\
$\dbsize$ & Size of the database in bits\\
$\dbblock$ & Size of one record of the database in bits\\
$\word$ & Size of one word of the database in bits\\
$\wnbr$ & Number of words per block\\
$\ind$ & Index of the record sought by \su\\
$\tp$ & Privacy level (tolerated number of colluding \db s)\\
$\kr$ & Number of responding \db s\\
$\vbr$ & Number of byzantine \db s\\
\noalign{\smallskip} \hline \noalign{\smallskip}
\end{tabular}
}
\end{table}

\noindent {\bf Private Information Retrieval ($\boldsymbol \pir$):} \pir~allows a user to retrieve a data item of its choice from a database, while preventing the server owning the database from gaining information on the identity of the item being retrieved~\cite{beimel2001information}. One trivial solution to this problem is to make the server send an entire copy of the database to the querying user. Obviously, this is a very inefficient solution to the \pir~problem as its communication complexity may be prohibitively large. However, it is considered as the only protocol that can provide information-theoretic privacy, i.e. perfect privacy, to the user's query in single-server setting. There are two main classes of \pir~protocols according to their privacy level: information-theoretic \pir~(\itpir) and computational \pir~(\cpir).
\begin{itemize}[leftmargin=*]
\item \noindent {\em Information-theoretic or multi-server \pir:} It guarantees information-theoretic privacy to the user, i.e. privacy against computationally unbounded servers. This could be achieved efficiently only if the database is replicated at $k \geq 2$ non-communicating servers~\cite{chor1998private,goldberg2007improving}. The main idea behind these protocols consists on decomposing each user's query into several sub-queries to prevent leaking any information about the user's intent.
\item \noindent {\em Computational or single-server \pir:} It guarantees privacy against computationally bounded server(s). In other words, a server cannot get any information about the identity of the item retrieved by the user unless it solves a certain computationally hard problem (e.g. prime factorization of large numbers), which is common in modern cryptography. Thus, they offer weaker privacy than their \itpir~counterparts~\cite{trostle2010efficient,melchor2008fast}.
\end{itemize}
\noindent {\bf Shamir Secret Sharing:} This is a concept introduced by Shamir et al.~\cite{shamir1979share} to allow a secret holder to divide its secret \secret~into \ns~shares $\secret_1, \cdots, \secret_{\ns}$ and distribute these shares to \ns~parties. In $(\tp,\ns)$-Shamir secret sharing, where $\tp < \ns$, if \tp~or fewer combine their shares, they learn no information about \secret. However, if more than \tp~come together, they can easily recover \secret. Given a secret \secret~chosen arbitrarily form a finite field, the $(\tp,\ns)$-Shamir secret sharing scheme works as follows: the secret holder chooses \ns~arbitrary non-zero distinct elements $\alpha_1,\cdots,\alpha_{\ns} \in \mathbb{F}$. Then, it selects \tp~elements $\sigma_1,\cdots,\sigma_{\tp}\in \mathbb{F}$ uniformly at random. Finally, the secret holder constructs the polynomial $f(x) = \sigma_0 + \sigma_1 x + \sigma_2 x^2 + \cdots + \sigma_t x^t$, where $\sigma_0 = \secret$. The \ns~shares $\secret_1, \cdots, \secret_{\ns}$, that are given to each party, are $(\alpha_1,f(\alpha_1)),\cdots,(\alpha_{\ns},f(\alpha_{\ns}))$. Any $\tp+1$ or more parties can recover the polynomial $f$ using Lagrange interpolation and thus they can reconstruct the secret $\secret = f(0)$. However, \tp~or less parties can learn nothing about \secret. In other words, if $\tp+1$ shares of \secret~are available then \secret~can be easily recovered.

\subsection{System Model and Security Definitions}
We consider a database-driven \crn~that contains \ns~\db s, where $\ns\geq 2$, and a \su~registered to these \db s to learn spectrum availability information in its vicinity. We assume that these \db s share the same content and that they are synchronized as mandated by PAWS standard~\cite{chen2015protocol}. We also assume that \db s may collude in order to infer \su's location. In the following, we present our security definitions.
\begin{definition}\label{def:byzantine}
\textbf{\em Byzantine} {\boldsymbol \db}{\bf:} This is a faulty \db~that runs but produces incorrect answers, possibly chosen maliciously or computed in error. This might be due to a corrupted or obsolete copy of the database caused by a synchronization problem with the other \db s.
\end{definition}

\begin{definition}\label{def:t-private}
\textbf{\bm\tp{\em -private}} {\boldsymbol\pir}{\bf:} The privacy of the query is information-theoretically protected, even if up to \tp~of the \ns~\db s collude, where $0 <\tp < \ns$.
\end{definition}

\begin{definition}\label{def:v-byzantine-robustness}
\textbf{$\bm\vbr${\em -Byzantine-robust}} {\boldsymbol\pir}{\bf:} Even if \vbr~of the responding \db s are {\em Byzantine}, \su~can reconstruct the correct database item, and determine which of the \db s provided incorrect response.
\end{definition}

\begin{definition}\label{def:k-robustness}
\textbf{{\em $\bm\kr$-out-of-$\bm\ns$}} {\boldsymbol\pir}{\bf:} \su~can reconstruct the correct record if it receives at least \kr-out-of-\ns~responses, $2\leq\kr\leq\ns$.
\end{definition}

\begin{definition}\label{def:robustness}
\textbf{{\em Robust}} {\boldsymbol\pir}{\bf:} It can deal with \db s that do not respond to \su's queries and allows \su~to reconstruct the correct output of the queries in this situation.
\end{definition}

\begin{definition}\label{def:tau-independence}
\textbf{$\bm\tau${\em -independent}} {\boldsymbol\pir}{\bf:} The content of the database itself is information theoretically protected from the coalition of up to $\tau$ \db s, where $0 \leq \tau < k-\tp$.
\end{definition}

\section{Proposed Approaches}
In the proposed approaches, we tailor multi-server \pir~to the context of multi-\db~\crn s. We start by illustrating the structure of the spectrum database that we consider. Then, we give several approaches, each adapts a multi-server \pir~protocol with different security, performance properties, and use cases.
We model the content of each \db~as an $\dbrow \times \wnbr$ matrix \dbmatrix~of size $\dbsize$ bits, where $\wnbr$ is the number of words of size \word~in each record/block of the database and $\dbrow$ is the number of records in the database, i.e. $\dbrow = \dbsize/\dbblock$, where $\dbblock=\wnbr\times\word$ is the block size in bits. The $k^{th}$ row of \dbmatrix~is the $k^{th}$ record of the database. 
\[ \dbmatrix = 
\begin{bmatrix}
    \word_{11} & \word_{12}  & \dots  & \word_{1\wnbr} \\
    \word_{21} & \word_{22}  & \dots  & \word_{2\wnbr} \\
    \vdots & \vdots  & \ddots & \vdots \\
    \word_{\dbrow 1} & \word_{\dbrow 2}  & \dots  & \word_{\dbrow\wnbr}
\end{bmatrix}
\]
We further assume that each row of the database corresponds to a unique combination of the tuple $(\x,\y,\chr,\ts)$, where \x~and \y~represent one location's latitude and longitude, respectively, \chr~is a channel number, and \ts~is a time-stamp. We also assume that \su s can associate their location information with the index \ind~of the corresponding record of interest in the database using some inverted index technique that is agreed upon with \db s. An \su~that wishes to retrieve record $\dbmatrix_\ind$ without any privacy consideration can simply send to \db~a row vector $\bm e_\ind$ consisting of all zeros except at position \ind~where it has the value $1$. Upon receiving $\bm e_\ind$, \db~multiplies it with \dbmatrix~and sends record $\dbmatrix_\ind$ back to \su~as we illustrate below: 
\[\begin{bmatrix}
    0 & \dots  & 0  & 1 & 0 & \dots & 0 \\
\end{bmatrix} 
\begin{bmatrix}
    \word_{11} & \word_{12}  & \dots  & \word_{1\wnbr} \\
    \word_{21} & \word_{22}  & \dots  & \word_{2\wnbr} \\
    \vdots & \vdots  & \ddots & \vdots \\
    \word_{\dbrow 1} & \word_{\dbrow 2}  & \dots  & \word_{\dbrow\wnbr}
\end{bmatrix}\]
\[= 
\begin{bmatrix}
    \word_{\ind 1} & \word_{\ind 2}  & \dots  & \word_{\ind\wnbr} \\
\end{bmatrix}\]
This trivial approach makes it easy for \db s to learn \su's location from the vector $\bm e_\ind$ as \dbmatrix~is indexed based on location. In the following we present two approaches that try to hide the content of $\bm e_\ind$ from \db s, and thus preserve \su's location privacy. The approaches present a tradeoff between efficiency, and some additional security features.

\subsection{Location Privacy with Chor (\chorScheme)}

\begin{figure}[h!]
\centering
\includegraphics[scale=0.33]{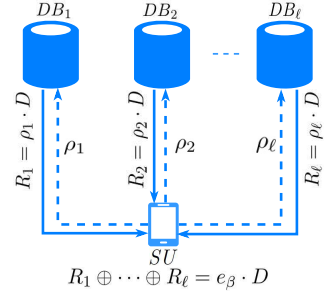} 
\caption{Main steps of \chorScheme~Algorithm}
\label{fig:chor}
\end{figure}

Our first approach, termed \chorScheme, harnesses the simple and efficient \itpir~protocol proposed by Chor et al.~\cite{chor1998private}. We describe the different steps of \chorScheme~in Algorithm~\ref{alg:chor} and highlight these steps in \figurename~\ref{fig:chor}. Elements of \dbmatrix~in this scheme belong to $GF(2)$, i.e. $\word = 1$ bit and $\dbblock = \wnbr$. 

\begin{algorithm}[h!]
\caption{$\dbmatrix_\ind\gets$\chorScheme(\ns, \dbrow, \dbblock)}\label{alg:chor}
\begin{algorithmic}[1]

\Statex \pmb \su 
\State $\ind \gets InvIndex(\x,\y,\chr,\ts)$
\State Sets standard basis vector $\bm{e}_\ind \gets \overrightarrow{1}_\ind \in \mathbb{Z}^\dbrow$
\State Generates $\bitstr_1, \cdots, \bitstr_{\ns-1} \in_R GF(2)^\dbrow$
\State $\bitstr_{\ns} \gets \bitstr_1 \oplus \cdots \oplus \bm{e}_\ind$
\State Sends $\bitstr_i$ to $\db_i$, for $1 \leq i \leq \ns$
\hspace{20pt}\algrule
\Statex {\bf Each $\pmb\db_i$}
\State Receives $\bitstr_i = \bitstr_{i1}\cdots\bitstr_{i\dbrow} \in \{0,1\}^\dbrow$
\State $\result_i \gets \bigoplus\limits_{\substack{1\leq j\leq \dbrow \\ \bitstr_{ij}=1}}\dbmatrix_j$, $\dbmatrix_j$ is the $j^{th}$ block of \dbmatrix
\State Sends $\result_i$ to \su
\hspace{20pt}\algrule
\Statex \pmb \su
\State Receives $\result_1,\cdots,\result_{\ns}$
\State $\dbmatrix_{\ind} \gets \result_1 \oplus\cdots\oplus \result_{\ns}$
\end{algorithmic}
\end{algorithm}
 In \chorScheme, \su~starts by invoking the inverted index subroutine $InvIndex(\x,\y,\chr,\ts)$ which takes as input the coordinates of the user, its channel of interest, and a time-stamp and returns a value \ind. This value corresponds to the index of the record $\dbmatrix_\ind$ of \dbmatrix~that \su~is interested in. \su~then constructs $\bm{e}_\ind$, which is a standard basis vector $\bm{\overrightarrow{1}_\ind} \in \mathbb{Z}^\dbrow$ having $0$ everywhere except at position $\ind$ which has the value $1$ as we discussed previously. \su~also picks $\ns-1$ \dbrow-bit binary strings $\bitstr_1, \cdots, \bitstr_{\ns-1}$ uniformly at random from $GF(2)^\dbrow$, and computes $\bitstr_{\ns} = \bitstr_1 \oplus \cdots \oplus \bm{e}_\ind$. Finally, \su~sends $\bitstr_i$ to $\db_i$, for $1 \leq i \leq \ns$. Upon receiving the bit-string $\bitstr_i = \bitstr_{i1} \oplus \cdots \bitstr_{i\dbrow}$ of length $\dbrow$, $\db_i$ computes $\result_i = \bitstr_i \cdot \dbmatrix$, which could be seen also as the XOR of those blocks $\dbmatrix_j$ in $\dbmatrix$ for which the $j^{th}$ bit of $\bitstr_i$ is $1$, then sends $\result_i$ back to \su. \su~receives $\result_i$s from $\db_i$s, $1 \leq i \leq \ns$, and computes $\result_1 \oplus \cdots \oplus \result_{\ns} = (\bitstr_1 \oplus \cdots \oplus \bitstr_{\ns}) \cdot \dbmatrix = \bm{e}_\ind \cdot \dbmatrix$, which is the $\ind^{th}$ block of the database that \su~is interested in, from which it can retrieve the spectrum availability information. 

\chorScheme~is very efficient thanks to its reliance on simple XOR operations only as we discuss in Section~\ref{sec:Performance}. It is also {$(\ns-1)$\em-private}, by Definition~\ref{def:t-private}, as collusion of up to $\ns-1$ \db s cannot enable them to learn $\bm{e}_\ind$, and consequently its location. In fact, only if \ns~\db s collude, then they will be able to learn $\bm{e}_\ind$ by simply XORing their $\{\bitstr_i\}_{i=1}^{\ns}$. However this approach suffers from two main drawbacks. First, it is not {\em robust} since even if one \db~fails to respond, \su~will not be able to recover $\dbmatrix_\ind$. Second, it is not {\em byzantine robust}; if one or more \db s return a wrong response, \su~will reconstruct a wrong block and also will not be able to recognize which \db~misbehaved so as not to rely on it for future queries. In Section~\ref{sec:goldberg} we discuss a second approach that improves on these two aspects but with some additional overhead.

\subsection{Location Privacy with Goldberg (\goldbergScheme)}\label{sec:goldberg}
\begin{figure}[h!]
\center
\includegraphics[scale=0.37]{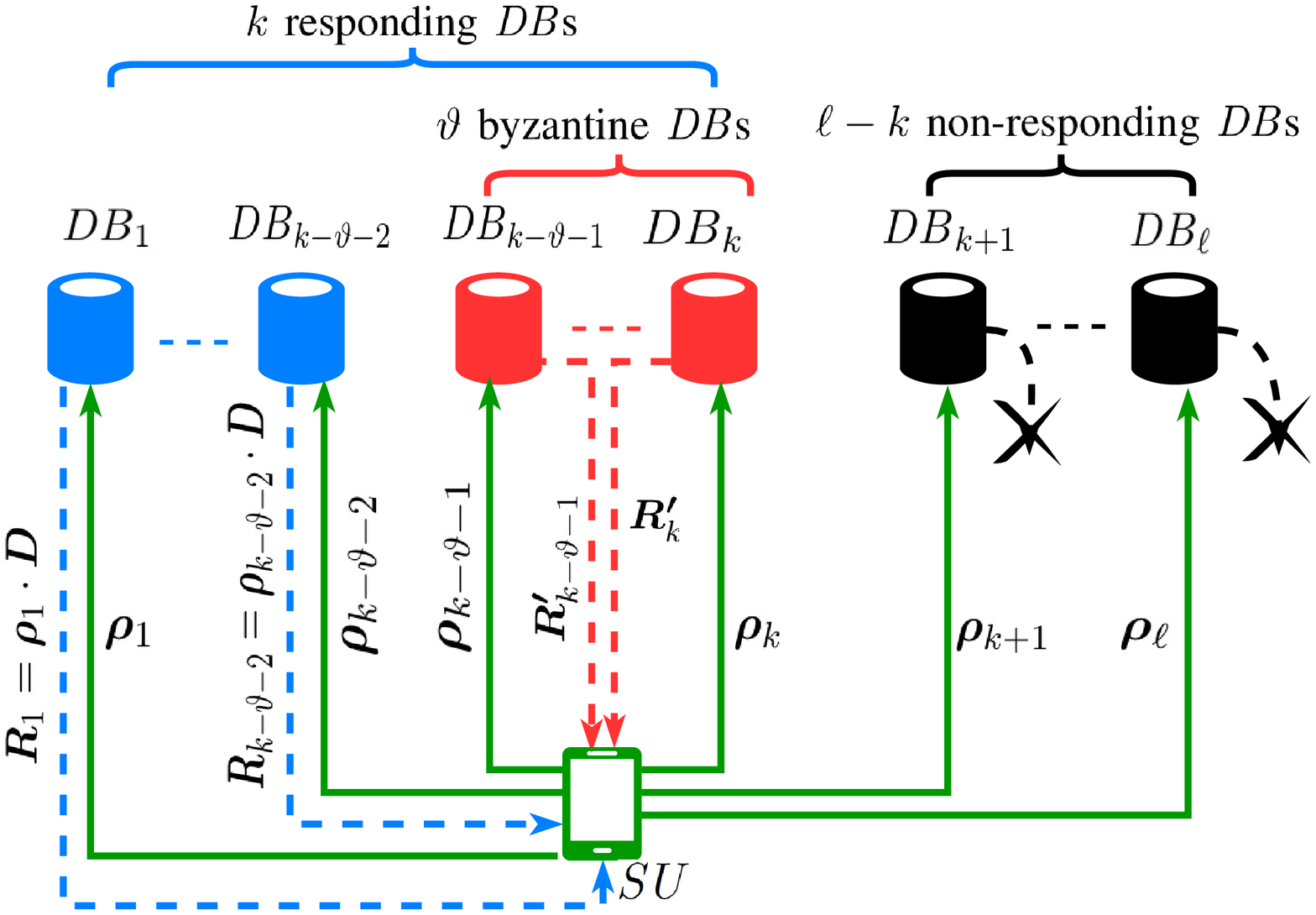} 
\caption{Illustration of \goldbergScheme}
\label{fig:pirGoldberg}
\end{figure}

Our second approach, termed \goldbergScheme, is based on Goldberg's \itpir~protocol~\cite{goldberg2007improving} which uses Shamir secret sharing to hide $\bm{e}_\ind$, i.e. \su's query. It is a modification of Chor's scheme~\cite{chor1998private} to achieve both {\em robustness} and {\em byzantine robustness}. Rather than working over $GF(2)$ (binary arithmetic), this scheme works over a larger field $\mathbb{F}$, where each element can represent $w$ bits. The database $\dbmatrix = (\word_{jk}) \in \mathbb{F}^{\dbrow \times \wnbr}$ in this scheme, is an $\dbrow \times \wnbr$ matrix of elements of $\mathbb{F} = GF(2^w)$. Each row represents one block of size $\dbblock$ bits, consisting of $\wnbr$ words of $\word$ bits each. Again, $\dbmatrix$ is replicated among $\ns$ databases $\db_i$. We summarize the main steps of \goldbergScheme~protocol in Algorithm~\ref{alg:goldberg} and illustrate them in \figurename~\ref{fig:pirGoldberg}.
  
To determine the index \ind~of the record that corresponds to its location, \su~starts by invoking the subroutine  $InvIndex(\x,\y,\chr,\ts)$ then constructs the standard basis vector $\bm{e}_\ind \in \mathbb{F}^r$ as explained earlier. \su~then uses $(\ns,\tp)$-Shamir secret sharing to divide the vector $\bm{e}_\ind$ into $\ns$ independent shares $(\alpha_1,,\bm{\dbshare}_1)\cdots,(\alpha_\ns,\bitstr_\ns)$ to ensure a $\tp${\em-private} \pir~protocol as in Definition~\ref{def:t-private}. That is, \su~chooses $\ns$ distinct non-zero elements $\alpha_i \in \mathbb{F}^*$ and creates $\dbrow$ random degree-$\tp$ polynomials $f_1,\cdots, f_\dbrow$ satisfying $f_j(0) = \bm{e}_\beta[j]$. \su~then sends to each $\db_i$ its share corresponding to the vector $\bitstr_i = \langle f_1(\alpha_i),\cdots, f_r(\alpha_i)\rangle$. Each $\db_i$ then computes the product $\bm{\result}_i = \bm{\dbshare}_i \cdot \dbmatrix = \langle\sum_j f_j(\alpha_i)\bm{\word}_{j1}, \cdots, \sum_j f_j(\alpha_i)\bm{\word}_{js}\rangle \in \mathbb{F}^s$ and sends $\bm{\result}_i$ to \su.

Some \db s may fail to respond to \su's query and only \kr-out-of-\ns~send their responses to \su. \su~collects \kr~responses from the \kr~responding \db s and tries to recover the record at index $\beta$ from the $\result_i$s by using the \textsc{EasyRecover()} subroutine from~\cite{goldberg2007improving} which uses Lagrange interpolation to recover $\dbmatrix_\ind$ from the secret shares $(\alpha_1,\result_1),\cdots,(\alpha_\kr,\result_{\kr})$. This is possible thanks to the use of $(\ns,\tp)$-Shamir secret sharing as long as $\kr > \tp$ and these \kr~\db s are honest. In fact, by the linearity property of Shamir secret sharing, since $\{(\alpha_i,\bm{\dbshare}_i)\}_{i=1}^{\ell}$ is a set of $(\ns,\tp)$-Shamir secret shares of $\bm{e}_\beta$, then $\{(\alpha_i, \result_i)\}_{i=1}^{\ell}$ will be also a set of $(\ns,\tp)$-Shamir secret shares of $\bm{e}_\beta \cdot \dbmatrix$, which is the $\beta^{th}$ block of the database. Thus, it is possible for \su~to reconstruct $\dbmatrix_\ind$ using Lagrange interpolation as explained in Section~\ref{sec:Prelim}, by relying only on the \kr~responses which makes \goldbergScheme~robust by Definition~\ref{def:robustness}. Also, the \textsc{EasyRecover} can detect the \db s that responded honestly, thus those that are byzantine as well, which should discourage \db s from misbehaving. More details about this subroutine could be found in~\cite{goldberg2007improving}.
\begin{algorithm}[h!]
\caption{$\dbmatrix_\ind\gets\goldbergScheme(\ns, \dbrow, \dbblock, \tp, \word)$}\label{alg:goldberg}

\begin{algorithmic}[1]
\Statex \pmb \su
\State $\ind \gets InvIndex(\x,\y,\chr,\ts)$
\State Sets standard basis vector $\bm{e}_\ind \gets \overrightarrow{1}_\ind \in \mathbb{Z}^\dbrow$
\State Chooses $\ell$ distinct $\alpha_1,\cdots,\alpha_\ns \in \mathbb{F}^*$
\State Creates $r$ random degree-$t$ polynomials $f_1,\cdots, f_r \in_R \mathbb{F}[x]$ s.t. $f_j(0) = \bm{e}_\beta[j]$' $\forall j \in [1,\cdots,\dbrow]$
\State $\bm{\dbshare}_i \gets \langle f_1(\alpha_i),\cdots, f_r(\alpha_i)\rangle$, $\forall i \in [1,\cdots,\ns]$
\State Sends $\bm{\dbshare}_i$ to $\db_i$, $\forall i \in [1,\cdots,\ns]$
\hspace{20pt}\algrule
\Statex {\bf Each honest $\pmb\db_i$}
\State Receives $\bm{\dbshare}_i$
\State $\bm{\result}_i \gets \bm{\rho}_i \cdot \bm{D} = \langle\sum_j f_j(\alpha_i)\bm{w}_{j1}, \cdots, \sum_j f_j(\alpha_i)\bm{w}_{js}\rangle$ \label{step:goldberg_multiplication}
\State Sends $\bm{\result}_i$ to \su
\hspace{20pt}\algrule
\Statex \pmb \su
\State Receives $\result_1,\cdots,\result_{\kr}$
\If{$\kr >\tp$}
\For{$c$ from $1$ to $s$}
\State $\result_{ic}\gets\result_{i}[c]$ $\forall i \in [1,\cdots,\kr]$
\State $\scode_c \gets \langle \result_{1c},\cdots,\result_{\kr c}\rangle$

\State $\dbmatrix_{\ind c}\gets \textsc{EasyRecover}(\tp,\word,[\alpha_1,\cdots,\alpha_\kr],\scode_c)$

\If{Recovery fails \textbf{and} $\vbr< \kr - \lfloor\sqrt{\kr \tp}\rfloor$}
\State $\scode_c \gets \langle \result_{1 c},\cdots,\result_{\kr c}\rangle$
\State $\dbmatrix_{\ind c}\gets \textsc{HardRecover}(\tp,\word,[\alpha_1,\cdots,\alpha_\kr],\scode_c)$
\EndIf
\EndFor
\EndIf

\end{algorithmic}
\end{algorithm}

Moreover, $\vbr$ \db s among the \kr~responding ones may even be {\em byzantine}, as in Definition~\ref{def:byzantine}, and  produce incorrect response. In that case, it would be impossible for \su~to simply rely on Lagrange interpolation to recover the correct responses. Since Shamir secret sharing is based on polynomial interpolation, the problem of recovering the response in the case of {\em byzantine} failures corresponds to noisy polynomial reconstruction, which is exactly the problem of decoding Reed-Solomon codes~\cite{devet2012optimally}. Thus, \su~would rather rely on error correction codes and more precisely on the Guruswami-Sudan list decoding~\cite{guruswami1998improved} algorithm which can correct $\vbr < \kr - \lfloor\sqrt{\kr \tp}\rfloor$ incorrect responses. In fact, the vector $\langle \bm{\result}_1[q], \bm{\result}_2[q],\cdots, \bm{\result}_\ell[q]\rangle$ is a Reed-Solomon code-word encoding the polynomial $g_q = \sum_j f_j \bm{\word}_{jq}$, and the client wishes to compute $g_q(0)$ for each $1 \leq q \leq \wnbr$ to recover all the \wnbr~words forming the record $\dbmatrix_\ind = \langle g_1(0),\cdots,g_\wnbr(0)\rangle$. This is done through the \textsc{HardRecover()} subroutine from~\cite{goldberg2007improving}. This makes \goldbergScheme~also \vbr{\em -Byzantine-robust}, by Definition~\ref{def:v-byzantine-robustness}, and solves the robustness issues that \chorScheme~suffers from, however, this comes at the cost of an additional overhead as we discuss in Section~\ref{sec:Performance}.

\begin{mycorollary}
 \chorScheme~and \goldbergScheme~directly inherit the security properties of Chor's~\cite{chor1998private} \pir~and Goldberg's~\cite{goldberg2007improving} \pir~respectively.
\end{mycorollary}

\subsection{Location Privacy of Mobile \su s Through Batching}

Thus far, we concerned only about non-mobile \su s that periodically submit an individual query to \db s to learn spectrum availability in their fixed location. However, things get more interesting with mobility. In fact, a mobile \su~will need to query \db s multiple times as its location changes. While the previous two approaches perform well for non-mobile \su s, they will incur a significant overhead on both \su~and \db s especially when \su~is moving at a relatively high speed, which will require a large number of \pir~queries.

Our third approach aims to protect the location privacy of mobile \su s while reducing the mobility-associated overhead. The idea is to exploit the fact that a mobile \su~usually has an a priori knowledge of its trajectory to make it query \db s for its current and future locations by batching these queries together instead of sending them separately. We achieve this by relying on the \itpir~protocol of Lueks et al.\cite{lueks2015sublinear} that extends the scheme of Goldberg~\cite{goldberg2007improving} to support batching of the queries using fast matrix multplication mechanisms inspired from batch codes~\cite{ishai2004batch}. We refer to this approach as \batchScheme~and we describe it in the following.

Each $\db_i$~that receives $\queries$ simultaneous queries $\bitstr_i^{(1)},\cdots,\bitstr_i^{(\queries)}$ from an \su~can process them using \goldbergScheme~by simply multiplying each query with \dbmatrix~as illustrated in Step~\ref{step:goldberg_multiplication} of Algorithm~\ref{alg:goldberg}. Alternatively, it can also group these queries into a matrix $\querymatrix_i$~of size $\queries\times\dbrow$, where each row $j$ corresponds to a query $\bitstr_i^{(j)}$, before computing the matrix product $\querymatrix_i\cdot\dbmatrix$. The careful reader will notice that this naive multiplication method would cost around $2qrs$ operations (including multiplications and additions) which can be prohibitively expensive especially for a large \dbmatrix~or \queries. This problem boils down to a fast matrix multiplication problem and therefore can benefit from fast matrix multiplication algorithms such as Strassen's~\cite{strassen1969gaussian}.

Strassen's algorithm consists on simply dividing both matrices $\querymatrix_i$~and \dbmatrix~ into four equally sized block matrices. Then instead of naively multiplying these submatrices, which will result in $8$ submatrix multiplications (fundamentally equivalent to  simple matrix multiplication), Strassen's algorithm creates linear combinations of blocks in a way that reduces the number of submatrix multiplications to $7$. The exact approach is then applied recursively to the multiplications of the submatrices of the previous step. This simple yet powerful matrix multiplication technique will significantly reduce the overhead for \db s and therefore the delay that \su s experience to learn spectrum availability while moving as illustrated in Section~\ref{sec:Performance}. 

A row $j$ in the resulting matrix, $\bm{\mathcal{R}_i} = \querymatrix_i\cdot\dbmatrix$, corresponds to $\db_i$'s response to the $j^{th}$ query. \su~will then recover the spectrum availability by combining same-index rows of the different $\bm{\mathcal{R}_i}$s as in \goldbergScheme.

\subsection{Location Privacy of \pu s}

As we mentioned earlier, in database-driven \crn s, \db s' content comprises operational information of \pu s which may be very sensitive in systems such as \sas~in the 3.5 GHz CBRS band where \pu s are military and governmental entities. The service providers use this operational data to feed their models and  populate the spectrum databases with availability information but do not share the \pu s' location information in response to \su s' queries. Therefore, \su s do not present a serious threat to \pu s privacy as opposed to the service providers which could be malicious, and could misuse \pu s' sensitive operational data.

In this subsection, we present another approach to take into account the privacy of these \pu s as well. For this we make use of another extension of the Goldberg~\pir~scheme known as $\tau${\em-independence}, to prevent \db s from learning the content of \dbmatrix~even if up to $\tau$ \db s collude to learn \dbmatrix~as defined in Definition~\ref{def:tau-independence}. This is achieved by making \pu s populate the \db s with spectrum availability information pertaining to their respective channels instead of the service providers, by secretly sharing each record they want to add, among the different service providers using Shamir secret sharing techniques, similar to how \su s secretly share their queries. That way, each service provider will not be able to decode this data, and only \su s which have access to the secret can retrieve the record by combining the different shares from the different DBs. This is motivated by the fact that \db s are expected to be populated by \pu s themselves as it is the case in LSA systems, or by a highly trusted independent entity, the ESC, as in \sas~systems. Therefore, whenever a \pu~or an ESC submits a \pu~activity record of index $j$ to \db s it will divide it into \wnbr~words $W_{j1}, \cdots, W_{j\wnbr}$ and distributes Shamir secret shares of every word among the $\ell$ \db s as reflected in Algorithm~\ref{alg:tau-indep-goldberg}. Each $\db_i$ will now have a different content $\dbmatrix^{(i)}$:

\[ \dbmatrix^{(i)} = 
\begin{bmatrix}
    \word^{(i)}_{11} & \word^{(i)}_{12}  & \dots  & \word^{(i)}_{1\wnbr} \\
    \word^{(i)}_{21} & \word^{(i)}_{22}  & \dots  & \word^{(i)}_{2\wnbr} \\
    \vdots & \vdots  & \ddots & \vdots \\
    \word^{(i)}_{\dbrow 1} & \word^{(i)}_{\dbrow 2}  & \dots  & \word^{(i)}_{\dbrow\wnbr}
\end{bmatrix}
\]

where $\{\word^{(i)}_{jc}\}_{1\leq i \leq \ns}$ form a $(\tau,\ns)$-Shamir secret sharing of word $W_{jc}$. 
This requires that the random values $\alpha_i$s, used to create Shamir secret shares as explained in Section~\ref{sec:blocks}, are shared beforehand among \su s and \pu s. This could be done by FCC during the registration phase, for instance, and must not be communicated to \db s.

\begin{algorithm}[h!]
\caption{$\dbmatrix_\ind\gets\tau$-$\goldbergScheme(\ns, \dbrow, \dbblock, \tp, \word)$}\label{alg:tau-indep-goldberg}

\begin{algorithmic}[1]
\Statex {\bf FCC}
\State Chooses $\ell$ distinct $\alpha_1,\cdots,\alpha_\ns \in \mathbb{F}^*$.
\State Shares these $\alpha_i$s only with \pu s and \su s.
\hspace{20pt}\algrule
\Statex \pmb \pu
\State Divides its activity record $j$ into $\wnbr$ words $W_{j1}, \cdots, W_{j\wnbr}$
\State Creates $\wnbr$ random degree-$\tau$ polynomials $g_{j1},\cdots,g_{j\wnbr}\in_R \mathbb{F}[x]$ s.t. $g_{jc}(0) = W_{jc}\; \forall c \in [1,\cdots,\wnbr]$
\State Sends $\word^{(i)}_{jc}\gets g_{jc}(\alpha_i)$ to $\db_i$, $\forall\:i\in[1,\cdots,\ns], \forall  c \in [1,\cdots,\wnbr]$
\State $\db_i$ adds $j^{th}$ record formed by $\word^{(i)}_{j1},\cdots,\word^{(i)}_{j\wnbr}$ to $\dbmatrix^{(i)}$
\hspace{20pt}\algrule
\Statex \pmb \su
\State $\ind \gets InvIndex(\x,\y,\chr,\ts)$
\State Sets standard basis vector $\bm{e}_\ind \gets \overrightarrow{1}_\ind \in \mathbb{Z}^\dbrow$
\State Creates $r$ random degree-$t$ polynomials $f_1,\cdots, f_r \in_R \mathbb{F}[x]$ s.t. $f_j(0) = \bm{e}_\beta[j]$ $\forall j \in [1,\cdots,\dbrow]$
\State $\bm{\dbshare}_i \gets \langle f_1(\alpha_i),\cdots, f_r(\alpha_i)\rangle$, $\forall i \in [1,\cdots,\ns]$
\State Sends $\bm{\dbshare}_i$ to $\db_i$, $\forall i \in [1,\cdots,\ns]$
\hspace{20pt}\algrule
\Statex {\bf Each honest $\pmb\db_i$}
\State Receives $\bm{\dbshare}_i$
\State $\bm{\result}_i \gets \bm{\rho}_i \cdot \dbmatrix^{(i)} = \langle\sum_j f_j(\alpha_i)\word^{(i)}_{j1}, \cdots, \sum_j f_j(\alpha_i)\word^{(i)}_{js}\rangle$ \label{step:new_goldberg_multiplication}
\State Sends $\bm{\result}_i$ to \su
\hspace{20pt}\algrule
\Statex \pmb \su
\State Receives $\result_1,\cdots,\result_{\kr}$
\If{$\kr >\tp+\tau$}
\For{$c$ from $1$ to $s$}
\State $\result_{ic}\gets\result_{i}[c]$ $\forall i \in [1,\cdots,\kr]$
\State $\scode_c \gets \langle \result_{1c},\cdots,\result_{\kr c}\rangle$
\State $\dbmatrix_{\ind c} \gets \textsc{EasyRecover}(\tp,\word,[\alpha_1,\cdots,\alpha_\kr],\scode_c)$
\If{Recovery fails \textbf{and} $\vbr< \kr - \lfloor\sqrt{\kr (\tp+\tau)}\rfloor$}
\State $\scode_c \gets \langle \result_{1c},\cdots,\result_{\kr c}\rangle$
\State $\dbmatrix_{\ind c}\gets \textsc{HardRecover}(\tp,\word,[\alpha_1,\cdots,\alpha_\kr],\scode_c)$
\EndIf
\EndFor
\EndIf

\end{algorithmic}
\end{algorithm}

This way, records revealing operational data of \pu s, which could be used by \db s to build knowledge of the activity of these \pu s and track them, are information-theoretically protected from \db s as long as no more than $\tau$ of these \db s collude. However, for this protocol to work, this condition must hold: $0<\tp\leq \tp+\tau < k \leq \ns$. While this extension of \goldbergScheme~should have no impact on the performance from \su s and \db s side as we show in Section~\ref{sec:Performance}, it has, however, an impact on the {\em t-privacy} of the protocol. In fact as the {\em $\tau$-independence} level, controlling how many \db s can collude to learn the record submitted by \pu, sought by \pu~increases, the maximum achievable {\em t-privacy level} will decrease since $\tp+\tau < k $ must always hold.

\subsection{Location Privacy of \su s in Partitioned-database \crn s}
In this section, we present another location privacy-preserving approach for \su s in the case where the spectrum database content is distributed among the different \db s instead of simply replicating it as in the previous approaches. This could be motivated by the fact that some database-driven \crn s may have multiple \db s covering different or slightly overlapping regions. It could also be a way to reduce cost by making each \db~manage a portion of the database. 

\begin{figure}[h!]
\centering
\includegraphics[width=0.35\textwidth]{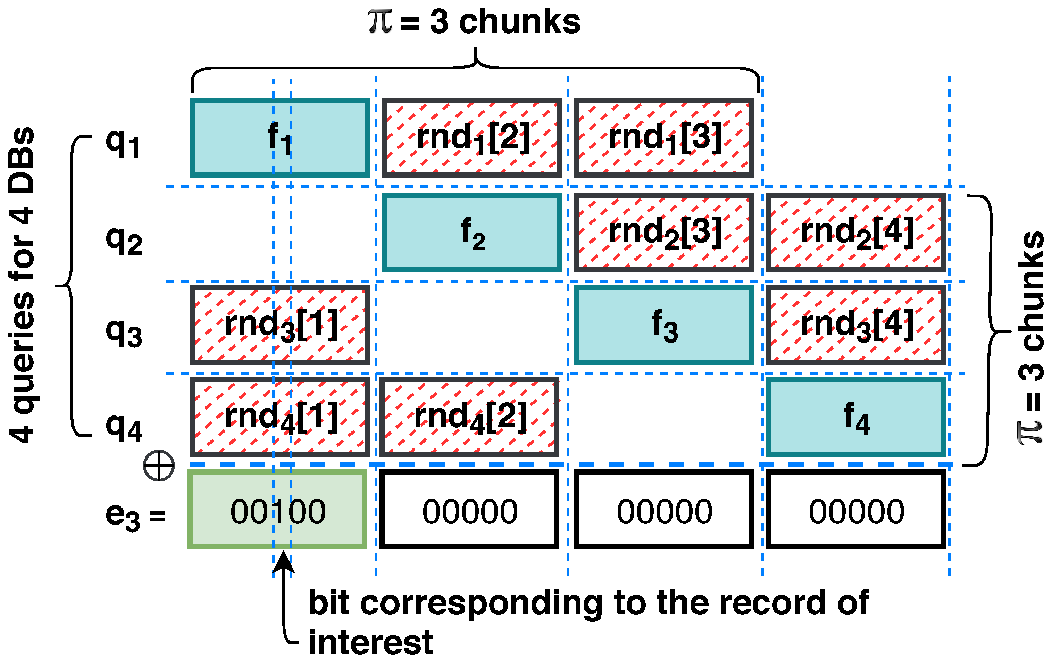} 
\caption{RAID-PIR~\cite{demmler2014raid}}
\label{fig:raid-pir}
 \vspace{-8pt}
\end{figure}

For that we rely on the RAID-PIR protocol due to Demmler et al.~\cite{demmler2014raid} which builds on Chor's scheme to reduce the communication overhead and the computation required at the server side. The idea here is very similar to that of Chor's but here the vector $\bm e_\ind$ is divided into \ns~chunks. Each query $q_i$ sent to $\db_i$ is divided into \redundancy~chunks as illustrated in Figure~\ref{fig:raid-pir}, where \redundancy~is a redundancy parameter that controls the minimum number of \db s that need to collude to recover the record $\dbmatrix_\ind$ with $2\leq\redundancy\leq\ns$. This parameter also controls the number of chunks in every query and how often the chunks overlap throughout these queries~\cite{demmler2014raid}.

The details of this approach are described in Algorithm~\ref{alg:raid}. To optimize the cost, \su~can use a pseudo random generator, $PRG$, to generate the $\redundancy-1$ chunks of $q_i$ as illustrated in Algorithm~\ref{alg:raid}. For that, \su~randomly generates \ns~seeds $s_1,\cdots,s_\ns$ of size $\kappa$ bits each, where $\kappa$ is the symmetric security parameter, and expands each seed $s_i$ into $\redundancy-1$ random chunks $rnd_i[j]$, using $PRG$, each of size $\frac{r}{\ns}$ as depicted in step~\ref{alg:expand} of Algorithm~\ref{alg:raid}. The first chunk of query $q_i$, denoted as $f_i$, is computed to cancel out the $\redundancy-1$ other $i^{th}$ chunks $rnd_i[j]$ of each of the other \db s, if applicable, and is obtained by xoring those $\redundancy-1$ chunks with the $i^{th}$ chunk of $\bm e_\ind$. Thanks to the use of the $PRG$, \su~does not need to send the whole query and needs only to send a compacted version of $q_i$, denoted as $q'_i$, composed of $f_i$ and the seed $s_i$, used to generate the other chunks of the full query $q_i$, to $\db_i$. Then, $\db_i$ will use the same pseudo-random generator, $PRG$, with the seed that it received to generate the full query $q_i$. Once $q_i$ recovered, $\db_i$ will construct its answer $\result_i$ by xoring the records in \dbmatrix~whose indices match those of the set bits in $q_i$. Finally, \su~needs only to xor the results from the different \db s to recover the $\ind^{th}$ record.

\begin{algorithm}[h!]
\caption{$\dbmatrix_\ind\gets RAID\mhyphen$\chorScheme(\ns, \dbrow, \dbblock)}\label{alg:raid}
\begin{algorithmic}[1]

\Statex \pmb \su 
\State $\ind \gets InvIndex(\x,\y,\chr,\ts)$
\State Sets standard basis vector $\bm{e}_\ind \gets \overrightarrow{1}_\ind \in \mathbb{Z}^\dbrow$
\State Picks \ns~seeds $s_i \in_R\{0,1\}^{\kappa}$
\State Expands $s_i$ to $\redundancy - 1$ chunks $ rnd_i[j] \gets PRG(s_i,j)\;\forall j\in[(i\;mod\;\ns)+1,(i+\redundancy-2\;mod\;\ns)+1]$, $\forall i\in[1,\ns]$\label{alg:expand}
\State $f_i\gets\bigoplus_j rnd_j[i]$, $j=(i-1\;mod\;\ns)+1, (i-2\;mod\;\ns+1),\cdots$
\State $f_i\gets {\bm e_\ind \oplus f_i}$ $\forall i\in[1,\ns]$
\State Sends $q'_i$ consisting of chunk $f_i$ and seed $s_i$ to $\db_i$

\hspace{20pt}\algrule
\Statex {\bf Each $\pmb\db_i$}
\State Expands its received $s_i$ as in Step~\ref{alg:expand} to get full query $q_i$

\State $\result_i \gets \bigoplus\limits_{\substack{1\leq j\leq \dbrow \\ q_{ij}=1}}\dbmatrix_j$, $\dbmatrix_j$ is the $j^{th}$ record of \dbmatrix
\State Sends $\result_i$ to \su
\hspace{20pt}\algrule
\Statex \pmb \su
\State Receives $\result_1,\cdots,\result_{\ns}$
\State $\dbmatrix_{\ind} \gets \result_1 \oplus\cdots\oplus \result_{\ns}$
\end{algorithmic}
\end{algorithm}

As the size of the query $q_i$ is just $\redundancy/\ns\cdot\dbrow$, each \db~now needs to store and process only $\redundancy/\ns\cdot\dbrow$ records of \dbmatrix~which will be beneficial to \db s especially if the number of these databases increases.

\section{Evaluation and Analysis}
\label{sec:Performance}
\subsection{Analytical Comparison }

We start by studying the proposed approaches' performance analytically and we compare them to existing approaches. For \goldbergScheme, we choose $\word = 8$ to simplify the cost of computations as in~\cite{devet2012optimally}; since in $GF(2^8)$, additions are XOR operations on bytes and multiplications are lookup operations into a $64$ KB table~\cite{devet2012optimally}. We summarize the system communication complexity and the computation incurred by both \db~and \su~and we illustrate the difference in architecture and privacy level of the different approaches in Table~\ref{tab:Table1}. As we mentioned earlier, existing research focuses on the single \db~setting. We compare the proposed approaches to existent techniques despite the difference of architecture to show the great benefits that multi-server \pir~brings in terms of performance and privacy as we discuss next. We briefly discuss these approaches in the following. 

Gao et al.~\cite{gao2013location} propose a \pir-based approach, termed \PrSpec, that relies on the \pir~scheme of Trostle et al.~\cite{trostle2010efficient} to defend against the new attack that they identify. This new attack exploits spectrum utilization pattern to localize \su s. Troja et al.~\cite{troja2014leveraging,troja2015efficient} propose two other \pir-based approaches that try to minimize the number of \pir~queries by either allowing \su s to share their availability information with other \su s~\cite{troja2014leveraging} or by exploiting trajectory information to make \su s retrieve information for their current and future positions in the same query~\cite{troja2015efficient}.

Despite their merit in providing location privacy to \su s these \pir-based approaches incur high overhead especially in terms of computation. This is due to the fact that they rely on \cpir~protocols to provide location privacy to \su s, which are known to suffer from expensive computational cost. In fact, answering an \su's query through a \cpir~protocol, requires \db~to process all of its records, otherwise \db~would learn that \su~is not interested in them and would then learn partial information about the record~$\dbmatrix_\ind$, and consequently \su's location. This makes the computational cost of most \cpir~based location preserving schemes linear on the database size from \db~side as we illustrate in Table~\ref{tab:Table1}. Now this is not exclusive to \cpir~protocols as even \itpir~protocols may require processing all the records to guarantee privacy, however, the main difference with \cpir~protocols is that the latter have a very large cost per bit in the database, usually involving expensive group operations like multiplication modulo a large modulus~\cite{aguilar2016xpir} as opposed to multi-server \itpir~protocols. This could be seen clearly in Table~\ref{tab:Table1} as both \chorScheme~and \goldbergScheme~require \db~to perform a very efficient XOR operation per bit of the database. The same applies to the overhead incurred by \su~which only performs XOR operations in both \chorScheme~and \goldbergScheme, while performing expensive modular multiplications and even exponentiations over large primes in the \cpir-based approaches. 

In terms of communication overhead, the proposed approaches incur a cost that is linear in the number of records $\dbrow$ and their size $\dbblock$. As an optimal choice of these parameters is usually $\dbrow = \dbblock = \sqrt{\dbsize}$~\cite{chor1998private,goldberg2007improving,devet2012optimally,aguilar2016xpir} then this cost could be seen as $\mathcal{O}(\sqrt{\dbsize\word})$ to retrieve a record of size $\sqrt{\dbsize\word}$ bits, which is a reasonable cost for an information theoretic privacy.

Moreover, as illustrated in Table~\ref{tab:Table1}, existent approaches fail to provide information theoretic privacy as the underlying security relies on computational \pir~schemes. The only approaches that provide information theoretic location privacy are \chorScheme, \goldbergScheme, and $RAID\mhyphen$\chorScheme~which are $(\ns-1)${\em -private}, {\em\tp-private}, and {\em ($\redundancy-1$)-private} respectively, by Definition~\ref{def:t-private}. It is worth mentioning that \PrSpec~\cite{gao2013location} relies on the well-known \cpir~of Trostle et al.~\cite{trostle2010efficient} representing the state-of-the-art in efficient \cpir. However, this \cpir~scheme has been broken~\cite{aguilar2016xpir,lepoint2015cryptanalysis}. Since the security of \PrSpec~follows that of Trostle et al.~\cite{trostle2010efficient} broken \cpir, then \PrSpec~fails to provide the privacy objective that it was designed for. However, we include it in our performance analysis for completeness. 

\subsection{Experimental Evaluation}
We further evaluate the performance of the proposed schemes experimentally to confirm the analytical observations.

\noindent \textbf{Hardware setting and configuration.} We have deployed the proposed approaches on GENI~\cite{Berman20145} cloud platform using the percy++ library~\cite{percy}. We have created $6$ virtual machines (VMs), each playing the role of a \db~and they all share the same copy of \dbmatrix. We deploy these GENI VMs in different locations in the US to count for the network delay and make our experiment closer to the real case scenario where spectrum service providers are located in different locations. These VMs are running Ubuntu $14.04$, each having $8$ GB of RAM, $15$ GB SSD, and $4$ vCPUs, Intel Xeon X5650 \@~$2.67$ GHz or Intel Xeon E5-2450\@~$2.10$ GHz. To assess the \su~overhead we use a Lenovo Yoga 3 Pro laptop with $8$ GB RAM running Ubuntu $16.10$ with an Intel Core m Processor 5Y70 CPU\@~$1.10$ GHz. The client laptop communicates with the remote VMs through ssh tunnels. We are also aware of the advances in \cpir~technology, and more precisely the fastest \cpir~protocols in the literature: XPIR which is proposed by Aguilar et al.\cite{aguilar2016xpir} and SealPIR due to Angel et al.~\cite{angel2018pir}. We include these protocols in our experiment to illustrate how multi-server \pir~performs against the best known \cpir~schemes if they are to be deployed in \crn s. We use the available implementation of these protocols provided in~\cite{Xpir} and~\cite{SealPIR} and we deploy their server components on a remote GENI VM while the client component is deployed on the Lenovo Yoga 3 Pro laptop.

\noindent \textbf{Dataset.} Spectrum service providers (e.g. Google, Microsoft, etc) offer graphical web interfaces and APIs to interact with their databases allowing to retrieve basic spectrum availability information for a user-specified location. Access to full data from real spectrum databases was not possible, thus, we generated random data for our experiment. The generated data consists of a matrix that models the content of the database, \dbmatrix, with a fixed block size $\dbblock = 560$ B while varying the number of records $\dbrow$. The value of $\dbblock$ is estimated based on the public raw data provided by FCC~\cite{cdbs} on a daily basis and which service providers use to populate their spectrum databases. 

\begin{table*}[ht!]

\centering  \caption{ Comparison with existent schemes} \label{tab:Table1}

\renewcommand{\arraystretch}{1.5}{
\resizebox{1\textwidth}{!}{%
\begin{tabular}{||c||c||c|c||c||c||}

\hline {\multirow{2}{*}{\textbf{Scheme} }}   & {\multirow{2}{*}{\textbf{Communication}}} & \multicolumn{2}{|c||}{\textbf{Computation}} &  \multicolumn{1}{|c||}{\multirow{2}{*}{\textbf{Setting}}} & {\multirow{2}{*}{\textbf{Privacy}}}\\ \cline{3-4}
 &  &  {$\boldsymbol  \db$} & {$ \boldsymbol \su$}& & \\ \hline

\hline \hline  \chorScheme &  $(\dbrow + \dbblock)\cdot \ns$ &  $\dbsize t_\oplus$ & $ (\dbrow+\dbblock)\cdot((\ns-1)\cdot t_\oplus)$ & \ns~\db s & $(\ns-1)${\em-private}\\

\hline  \goldbergScheme & $\dbrow\cdot\word\cdot\ns + \kr\cdot\dbblock$ & $(\dbsize/\word)\cdot t_\oplus$ & $\ns\cdot(\ns-1)\cdot\dbrow t_{\oplus} + 3\ns\cdot(\ns+1)t_{\oplus}$ & \ns~\db s& \tp{\em-private \ns-comp.-private} \\

\hline  $RAID\mhyphen$\chorScheme & $\dbrow+\ns\cdot\kappa+\ns\cdot\dbblock$ & $(\redundancy/\ns)\cdot \dbsize t_\oplus$ & $( \dbrow\cdot(\redundancy-1) + \dbblock\cdot(\ns-1)) t_\oplus$ & \ns~\db & $(\redundancy-1)${\em-private} \\

\hline  \hline \PrSpec~\cite{gao2013location} & $(2\sqrt{\dbrow}+3)\cdot\lceil\log p\rceil$  & $\mathcal{O}(\dbrow)\cdot Mulp$ & $4\sqrt{\dbrow}\cdot Mulp$& $1$ \db & {\em underlying \pir~ broken}\\
\hline  Troja et al~\cite{troja2015efficient} & $12\delta\cdot\dbblock$ & $\mathcal{O}(\dbsize)\cdot Mulp$ & $4\sqrt{\dbsize}\cdot Mulp$ & $1$ \db & {\em computationally-private}\\ 

\hline  Troja et al~\cite{troja2014leveraging} & $n_g\cdot \psi\cdot\log_2q + (2\sqrt{\dbsize}+3)\cdot\lceil\log p\rceil$ & $\mathcal{O}(\dbsize)\cdot Mulp$ & $n_g\cdot \psi\cdot(2Expp+Mulp)+4\sqrt{\dbsize}\cdot Mulp$& $1$ \db & {\em computationally-private}\\

\hline  XPIR~\cite{aguilar2016xpir} & $\mathcal{O}(Nd \sqrt[\leftroot{0}\uproot{2}d]{\dbsize})$ & $2d\cdot(\dbrow/\alpha)\cdot(\dbblock/\ell_0)\cdot Mulp$ & $d\cdot(\dbrow/\alpha)^{1/d}\cdot Enc + d\cdot\alpha\cdot\dbblock/\ell_0\cdot Dec$ & $1$ \db & {\em computationally-private}\\ \hline

\hline SealPIR~\cite{angel2018pir} & $\mathcal{O}(Nd \lceil\sqrt[\leftroot{0}\uproot{2}d]{\dbsize}/N\rceil)$ & $\mathcal{O}(d \sqrt[\leftroot{0}\uproot{2}d]{\dbsize})$ & $d\cdot\mathcal{E} +  (F^{d-1}+1)\cdot \mathcal{D}$ & $1$ \db & {\em computationally-private}\\ \hline
\end{tabular}}}

\flushleft{\scriptsize{\textbf{Variables:} $t_\oplus$ is the execution time of one XOR operation. $p$ is a large prime, and $Mulp$ and $Expp$ are the execution time of performing one modular multiplication, and one modular exponentiation respectively. $\psi$~denotes the number of bits that an \su~shares with other \su s in~\cite{troja2014leveraging}, $n_g$ is the number of \su s within a same group in~\cite{troja2014leveraging}. $\delta$ is the number of \db~segments in~\cite{troja2015efficient}. $d$ is the recursion level, $\alpha$ is the aggregation level, $\mathcal{C}$ is the Ring-LWE ciphertext size, $\lambda$ is the number of elements returned by \db, $F$ is the expansion factor of the underlying cryptosystem, $\ell_0$ is the number of bits absorbed in a cyphertext, all are used in~\cite{aguilar2016xpir}. $(Enc,Dec)$ are respectively the encryption and decryption cost for Ring-LWE cryptosystem used in~\cite{aguilar2016xpir}. $(\mathcal{E},\mathcal{D})$ are respectively the encryption and decryption cost for Fan-Vercauteren~\cite{fan2012somewhat} cryptosystem used in~\cite{angel2018pir}. $N$ is the query size bound in XPIR and SealPIR and is typically is typically 2048 or 4096 based on recommended security parameters.

}}
\end{table*}

\begin{figure}[h!]
\centering
\includegraphics[width=0.52\textwidth]{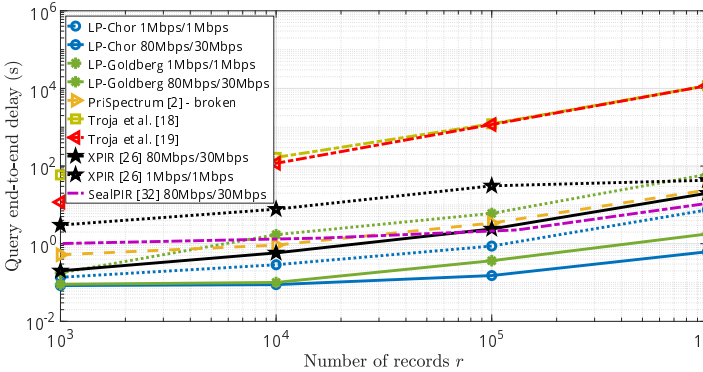} 
\caption{Query RTT of the different PIR-based approaches}
\label{fig:queryRTT}
 \vspace{-8pt}
\end{figure}

\noindent \textbf{Results and Comparison.} We first measure the query end-to-end delay of the proposed approaches and plot the results in \figurename~\ref{fig:queryRTT}. We also include the delay introduced by the existing schemes based on our estimation of the operations included in Table~\ref{tab:Table1}. The end-to-end delay that we measure takes into consideration the time needed by \su~to generate the query, the network delay, the time needed by \db~to process the query, and finally the time needed by \su~to extract the $\ind^{th}$ record of the database. We consider two different internet speed configurations in our experiment. We first rely on a high-speed internet connection of $80 Mbps$ on the download and $30 Mbps$ on the upload for all compared approaches. Then we use a low-speed internet connection of $1 Mbps$ on the upload and download to assess the impact of the bandwidth on \chorScheme~and \goldbergScheme, and also on XPIR as well.

\figurename~\ref{fig:queryRTT} shows that the proposed schemes perform much better than the existing approaches in terms of delay even with low-speed internet connection. They also perform better than the fastest existing \cpir~protocols XPIR and SealPIR. This shows the benefit of relying on multi-server \itpir~in multi-\db~\crn s. Also, and as expected, \chorScheme~scheme performs better than \goldbergScheme~thanks to its simplicity. As we will see later, \goldbergScheme~also incurs larger communication overhead than \chorScheme~as well. This could be acceptable knowing that \goldbergScheme~can handle collusion of up-to \ns~\db s, and is robust in the case of $(\ns-\kr)$ non-responding \db s, and \vbr~byzantine \db s, as opposed to \chorScheme. This means that \goldbergScheme~could be more suitable to real world scenario as failures and byzantine behaviors are common in reality. \figurename~\ref{fig:queryRTT} also shows that the network bandwidth has a significant impact on the end-to-end latency. This is due to the relatively large amount of data that needs to be exchanged during the execution of these protocols which requires higher internet speeds.

\begin{figure}[!h]
    \centering
    \subcaptionbox{\small SU Computation Overhead.\label{fig:suComuputation}}
    {\includegraphics[width=0.23\textwidth]{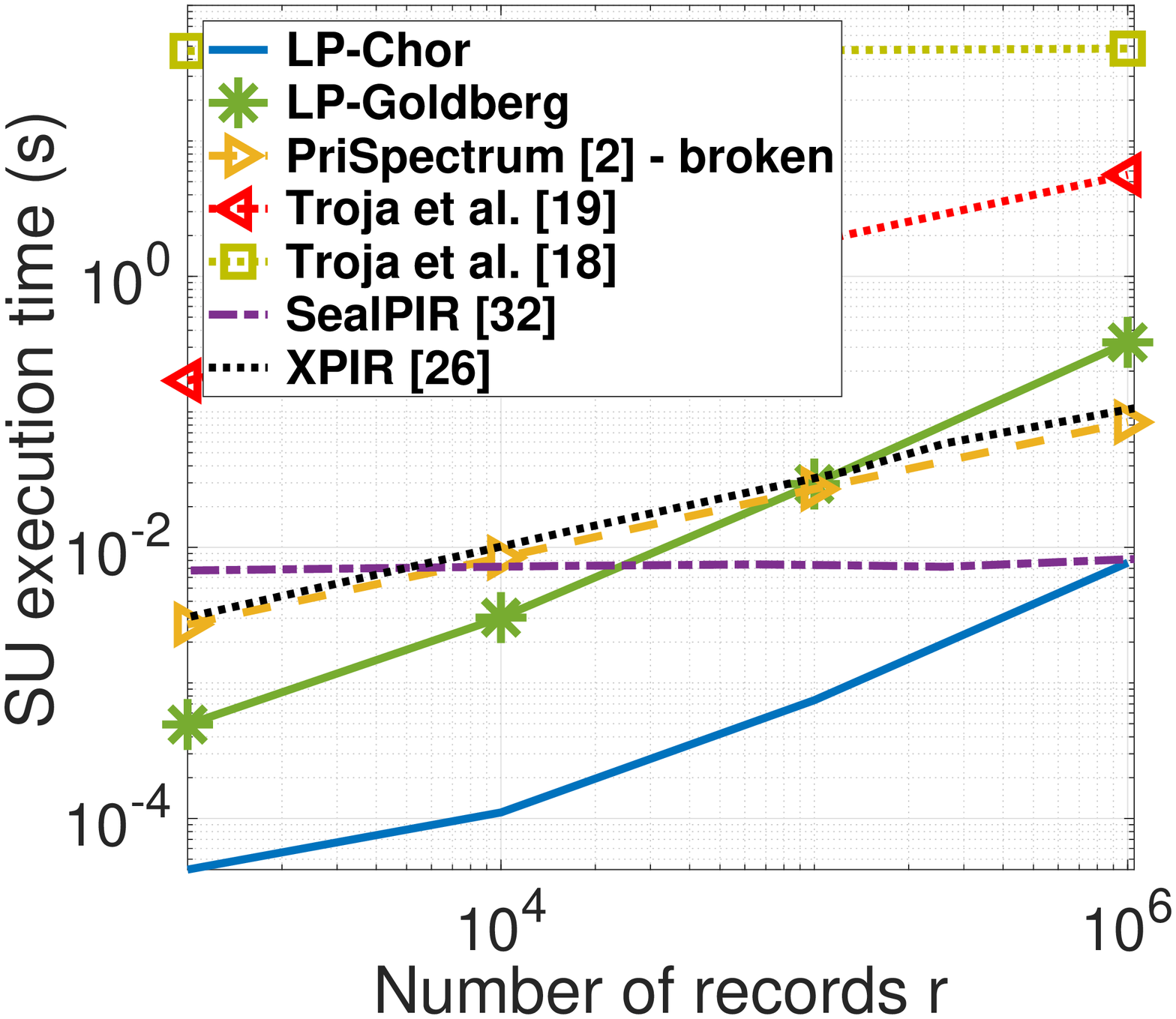}}\quad
    \subcaptionbox{\small DB Computation Overhead.\label{fig:dbComuputation}}
     {\includegraphics[width=0.23\textwidth]{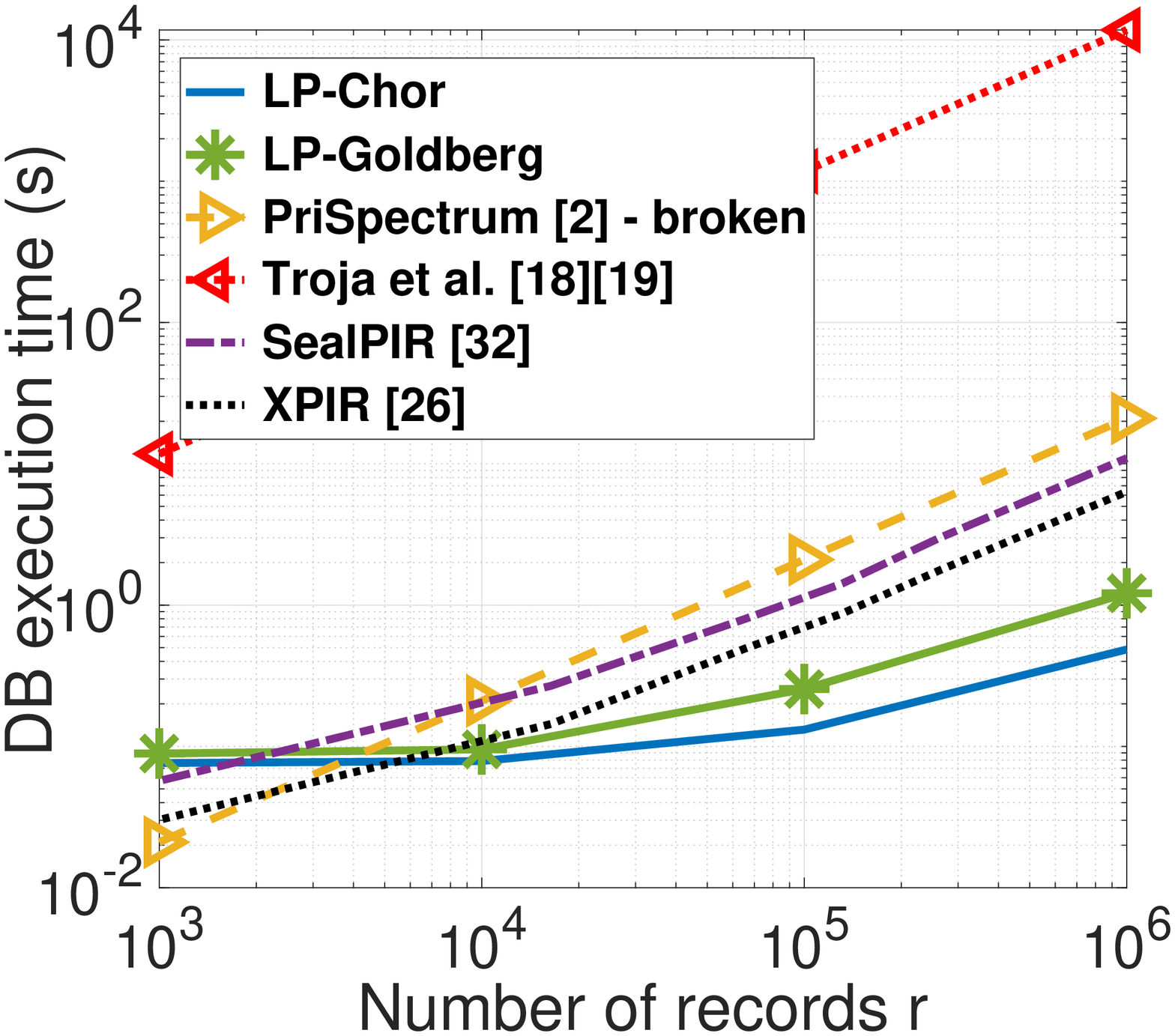}}
    \caption{Computation Comparison}
    \label{fig:comp}
    \vspace{-8pt}
\end{figure}

We also compare the computational complexity experienced by each \su~and \db~separately in the different approaches as shown in Table~\ref{tab:Table1}. We further illustrate this through experimentation and we plot the results in \figurename~\ref{fig:suComuputation}, which shows that the proposed schemes incur lower overhead on the \su~than the existing approaches. The same observation applies to the computation experienced by each \db~which again involves only efficient XOR operations in the proposed schemes. We illustrate this in \figurename~\ref{fig:dbComuputation}.

\begin{figure}[h!]
\center
\includegraphics[width=0.35\textwidth]{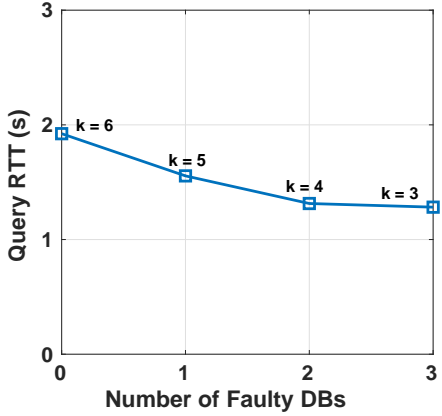} 
\caption{Impact of the number of faulty \db s on the query RTT.}
\label{fig:robustness}
 \vspace{-8pt}
\end{figure}

We also study the impact of non-responding \db s on the end-to-end delay experienced by the \su~in \goldbergScheme~as illustrated in \figurename~\ref{fig:robustness}. This Figure shows that as the number of faulty \db s increases, the end-to-end delay decreases since \su~needs to process fewer shares to recover the record $\dbmatrix_\ind$. As opposed to \chorScheme, in \goldbergScheme, \su~is still able to recover the record \ind~even if only \kr~out-of-\ns~\db s respond. Please recall also that our experiment was performed on resource constrained VMs to emulate \db s, however in reality, \db s should have much more powerful computational resources than those of the used VMs which will have a tremendous impact on further reducing the overhead of the proposed approaches.

\begin{figure}[!h]
    \centering
    \subcaptionbox{\small \su~Computation Overhead.\label{fig:privacy-level-su}}
    {\includegraphics[width=0.23\textwidth]{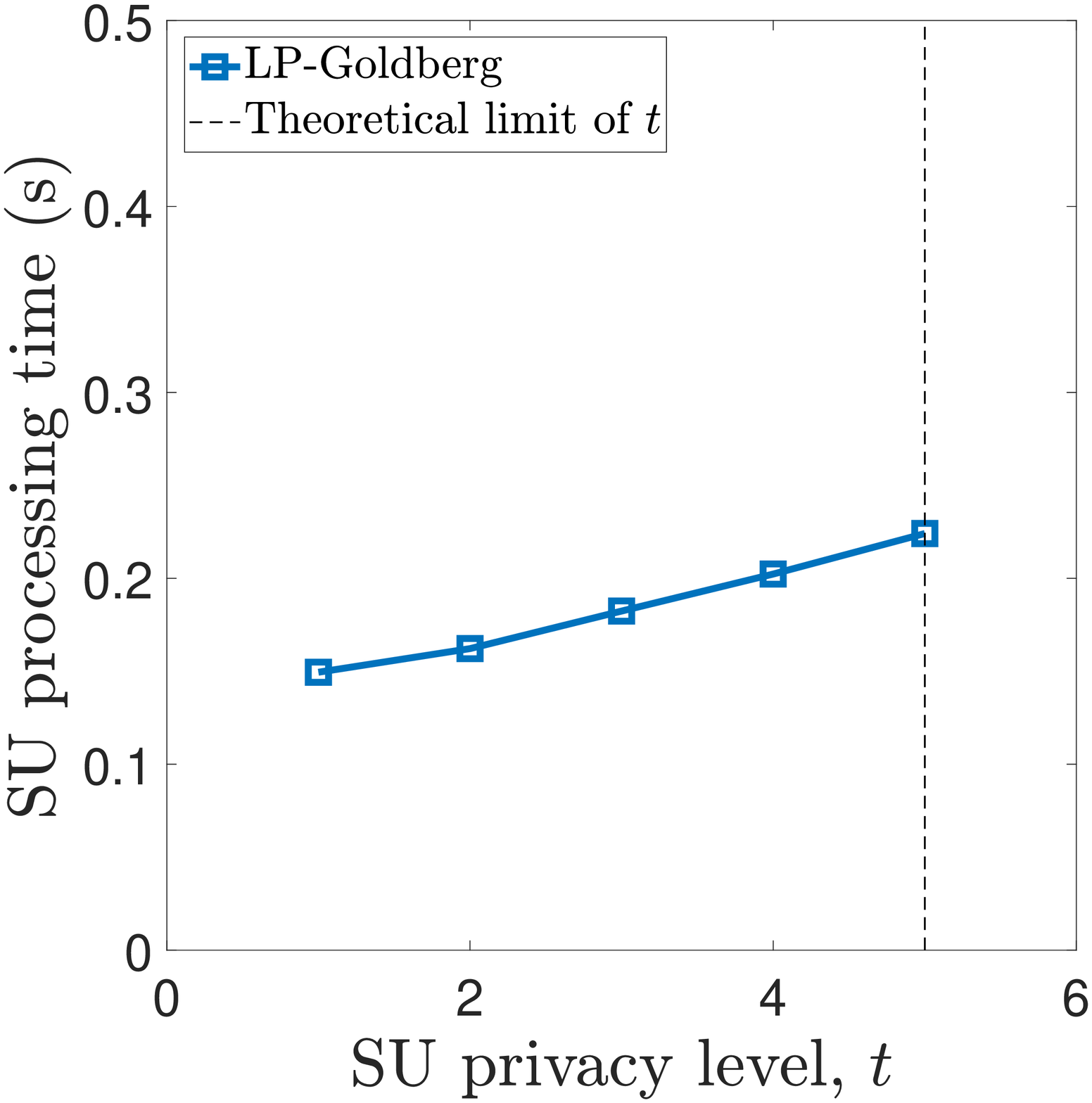}}\quad
    \subcaptionbox{\small \db~Computation Overhead.\label{fig:privacy-level-db}}
     {\includegraphics[width=0.23\textwidth]{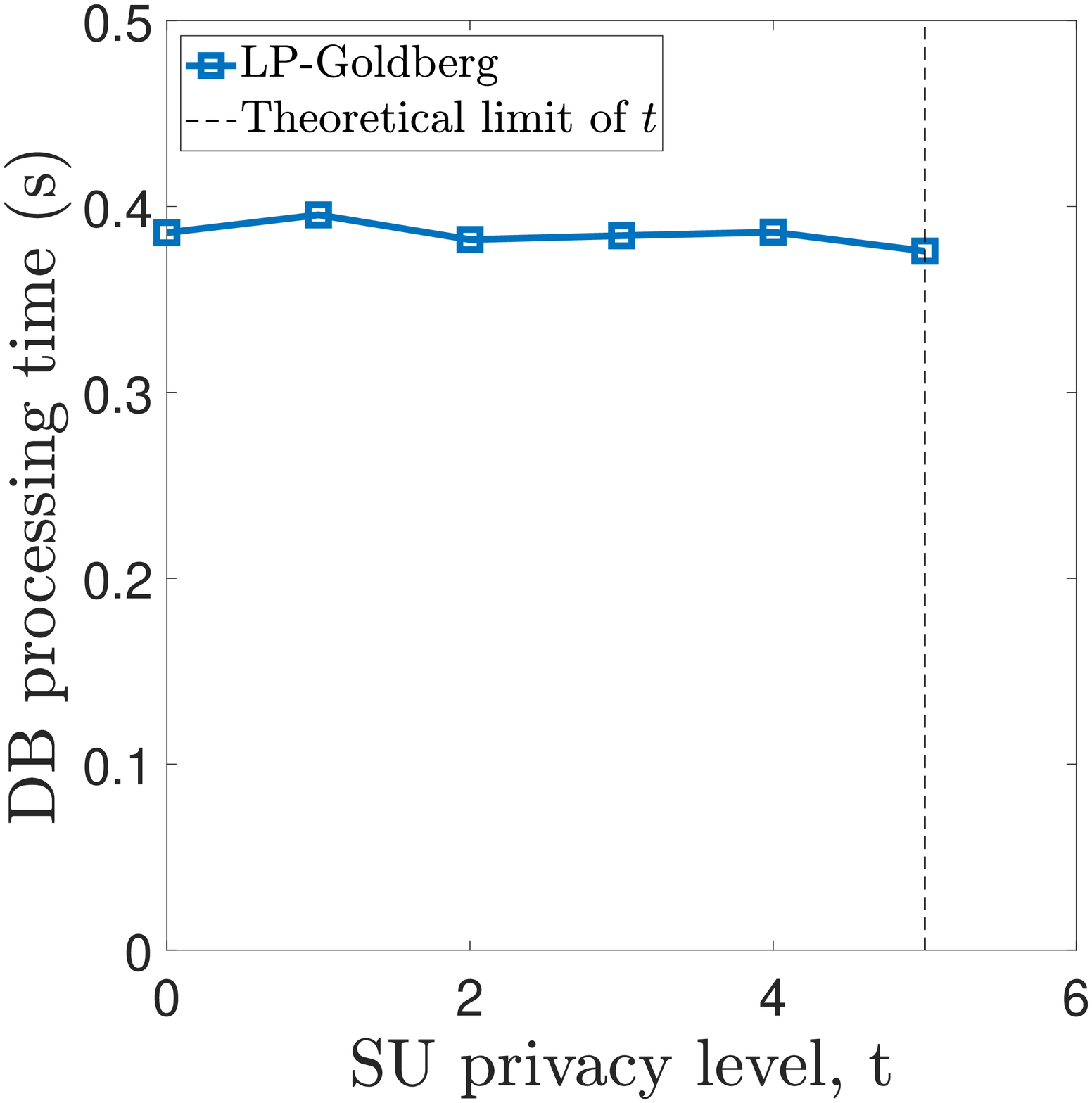}}
    \caption{Impact of increasing query privacy level, \tp}
    \label{fig:privacy-level}
    \vspace{-10pt}
\end{figure}
Figure~\ref{fig:privacy-level} illustrates the impact of \su's desired privacy level in \goldbergScheme~on the processing time incurred by both \su~and \db s. As expected, increasing the value of \tp, which controls the number of \db s that can collude without inferring the content of the query, should not have any impact on each \db~as they will always perform the same operations regardless of the privacy level. However, since the results sent by \db s could also be considered as a $(\tp,\ns)$-Shamir secret sharing of the retrieved record, when \tp~increases, then the number of secret shares required to recover the record increases which will result in more computation for the \su~when performing Lagrange interpolation over higher degree-\tp~polynomials.

\begin{figure}[h!]
\center
\includegraphics[width=0.53\textwidth]{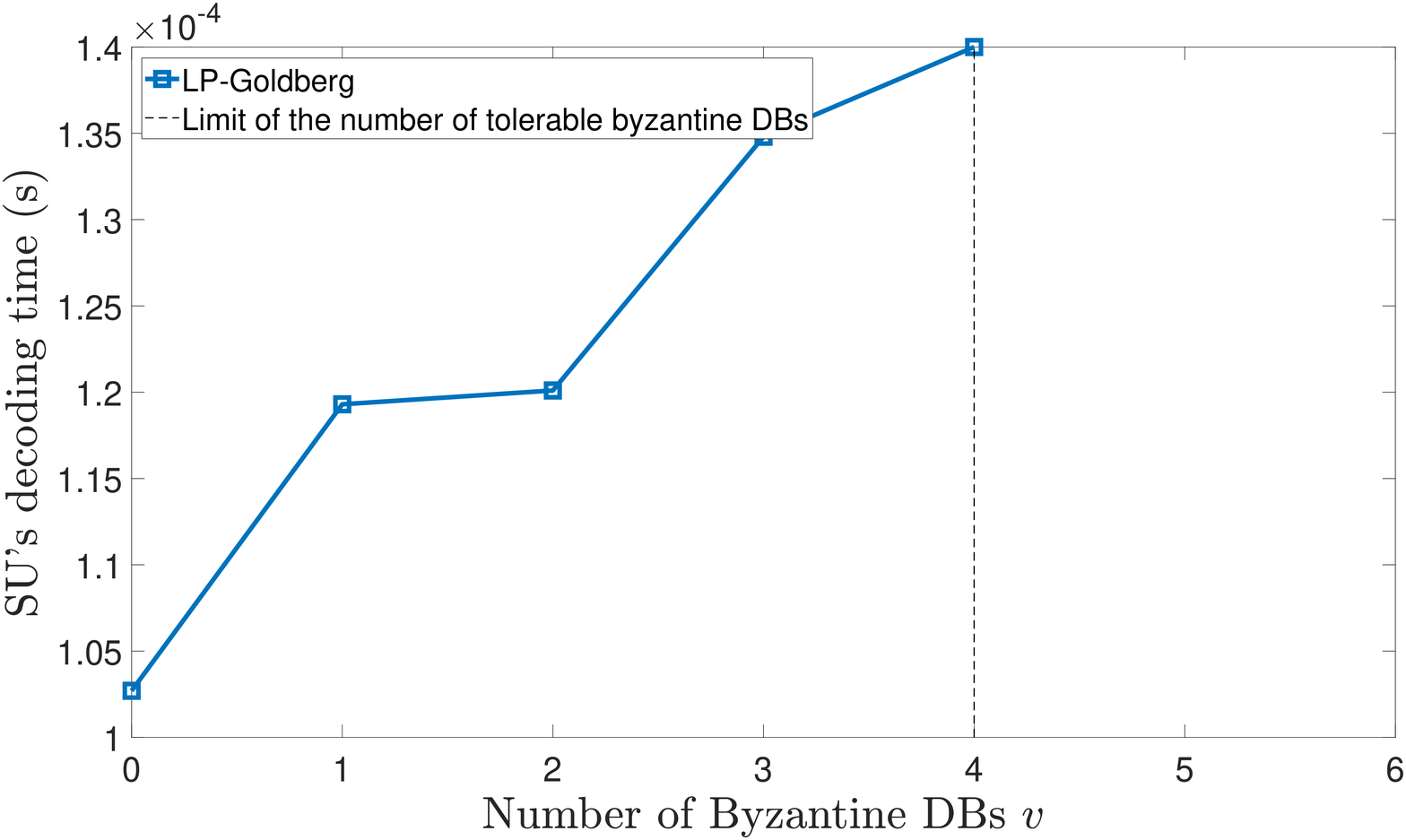} 
\caption{ Performance of \goldbergScheme~in the presence of byzantine \db s}
\label{fig:byzantine}
 \vspace{-10pt}
\end{figure}

We further study the impact of the number of byzantine \db s on the processing time on \su~side in \goldbergScheme~as depicted in Figure~\ref{fig:byzantine}. As expected, having more byzantine \db s will increase the complexity of decoding the different shares, that \su~receives from \db s, using the relatively expensive \textsc{HardRecover} subroutine from~\cite{goldberg2007improving}.

\begin{figure}[h!]
\center
\includegraphics[width=0.36\textwidth]{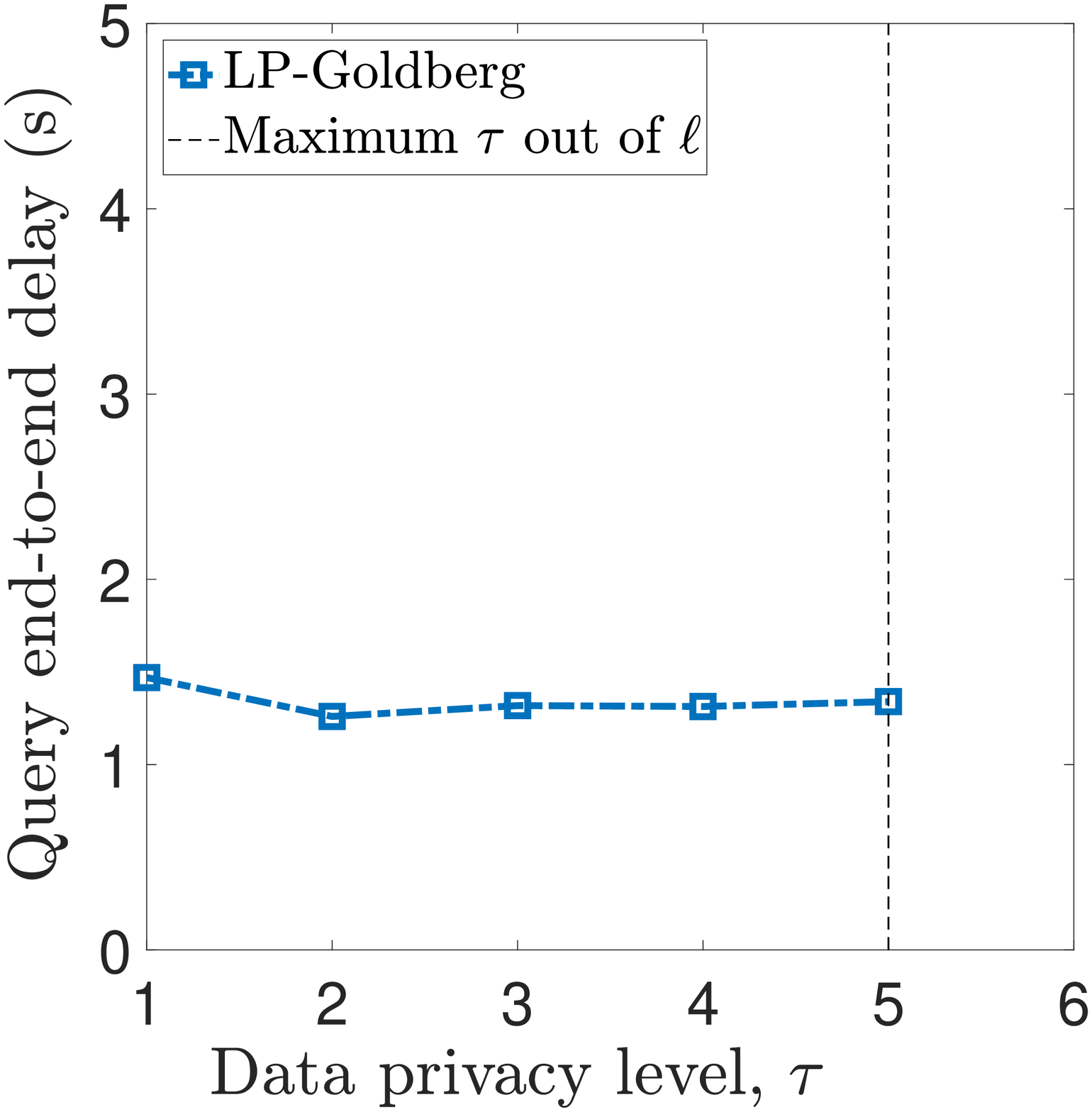} 
\caption{Performance of $\tau$-independent \goldbergScheme, with $k=\ns =6$ and $\tp+\tau<k$}
\label{fig:pu-privacy}
 \vspace{-10pt}
\end{figure}

As for $\tau$-\goldbergScheme, the $\tau$-{\em independence} extension will have no impact on the processing time of \db s and should also have no impact on \su s as long as $\tp+\tau$ is constant. This means that both \pu s and \su s will always seek the maximum privacy levels for their data and queries such that $\tp + \tau < k$. This is reflected in Figure~\ref{fig:pu-privacy}. However the processing time will be linear in $\tp+\tau$ similar to Figure~\ref{fig:privacy-level-su}.

\begin{figure}[h!]
\center
\includegraphics[width=0.53\textwidth]{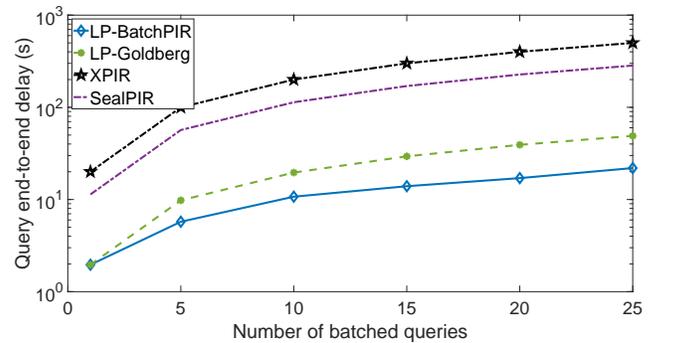} 
\caption{Query RTT for a moving \su}
\label{fig:queryRTT-mobiliy}
 \vspace{-10pt}
\end{figure}

As for the case of mobile \su s, we compare the performance of batching multiple queries for the future locations of a \su~to that of sending separate consecutive queries using \goldbergScheme, SealPIRand,and XPIR as depicted in Figure~\ref{fig:queryRTT-mobiliy}. Using batching mainly reduces the computation on \db s side and will reduce the end-to-end delay for answering the queries of the moving \su. 

We also demonstrate the benefit of relying on $RAID\mhyphen$\chorScheme~and partitioning the database content among \db s, instead of simply replicating it, on the \db s' side for several values of the redundancy parameter \redundancy. As expected, $\redundancy=2$ yields the best performance however it also offers the lowest level of resistance to collusion. Setting $\redundancy$ to be equal to \ns~will is equivalent to the original scheme \chorScheme~and will have the best performance. Therefore, $RAID\mhyphen$\chorScheme~offers a performance-privacy tradeoff that is controlled by the redundancy parameter $\redundancy$. 

\begin{figure}[h!]
\center
\includegraphics[width=0.36\textwidth]{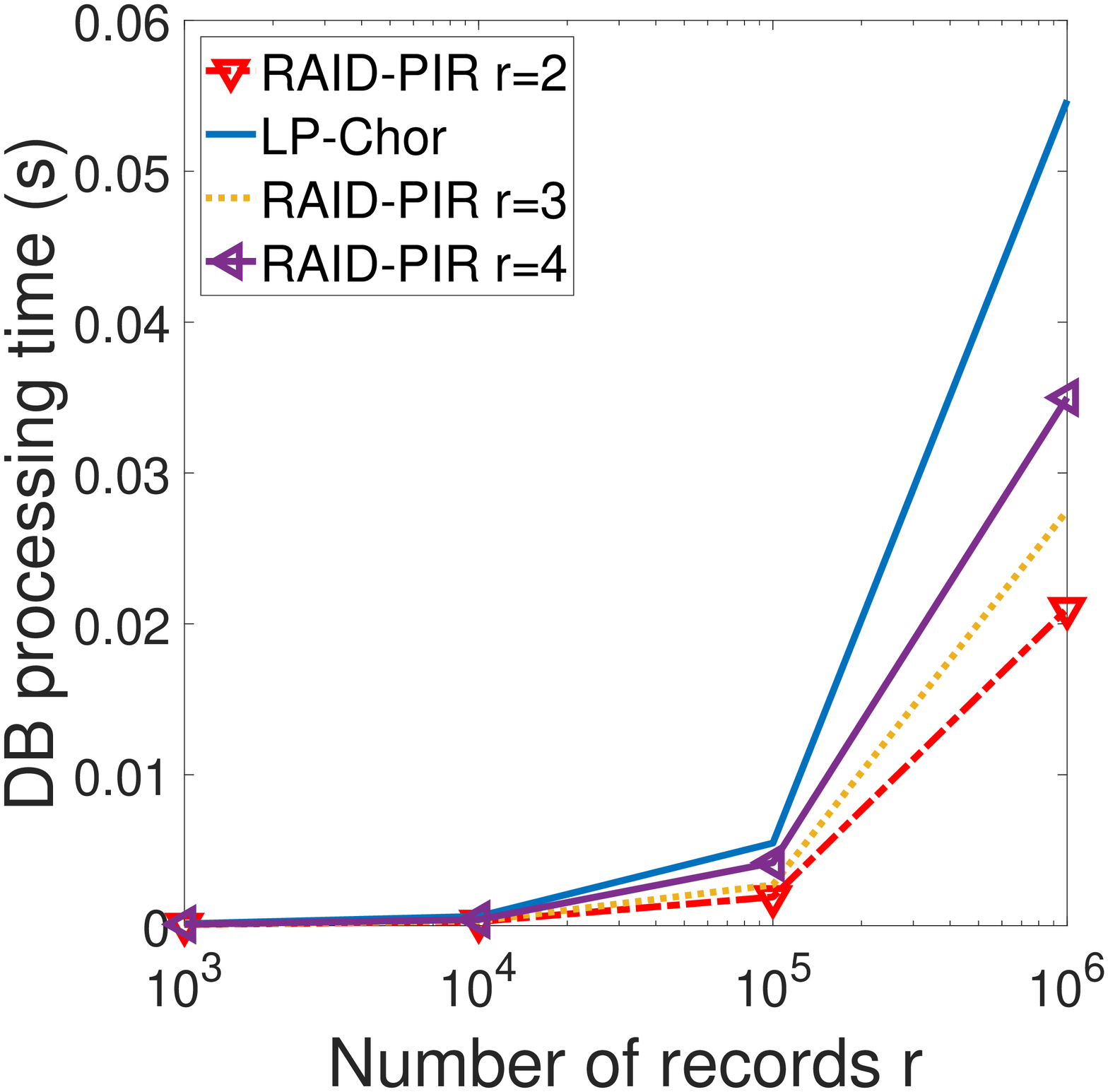} 
\caption{\db's processing time under $RAID\mhyphen$\chorScheme~compared to \chorScheme}
\label{fig:raid}
 \vspace{-10pt}
\end{figure}

\begin{figure}[h!]
\center
\includegraphics[width=0.53\textwidth]{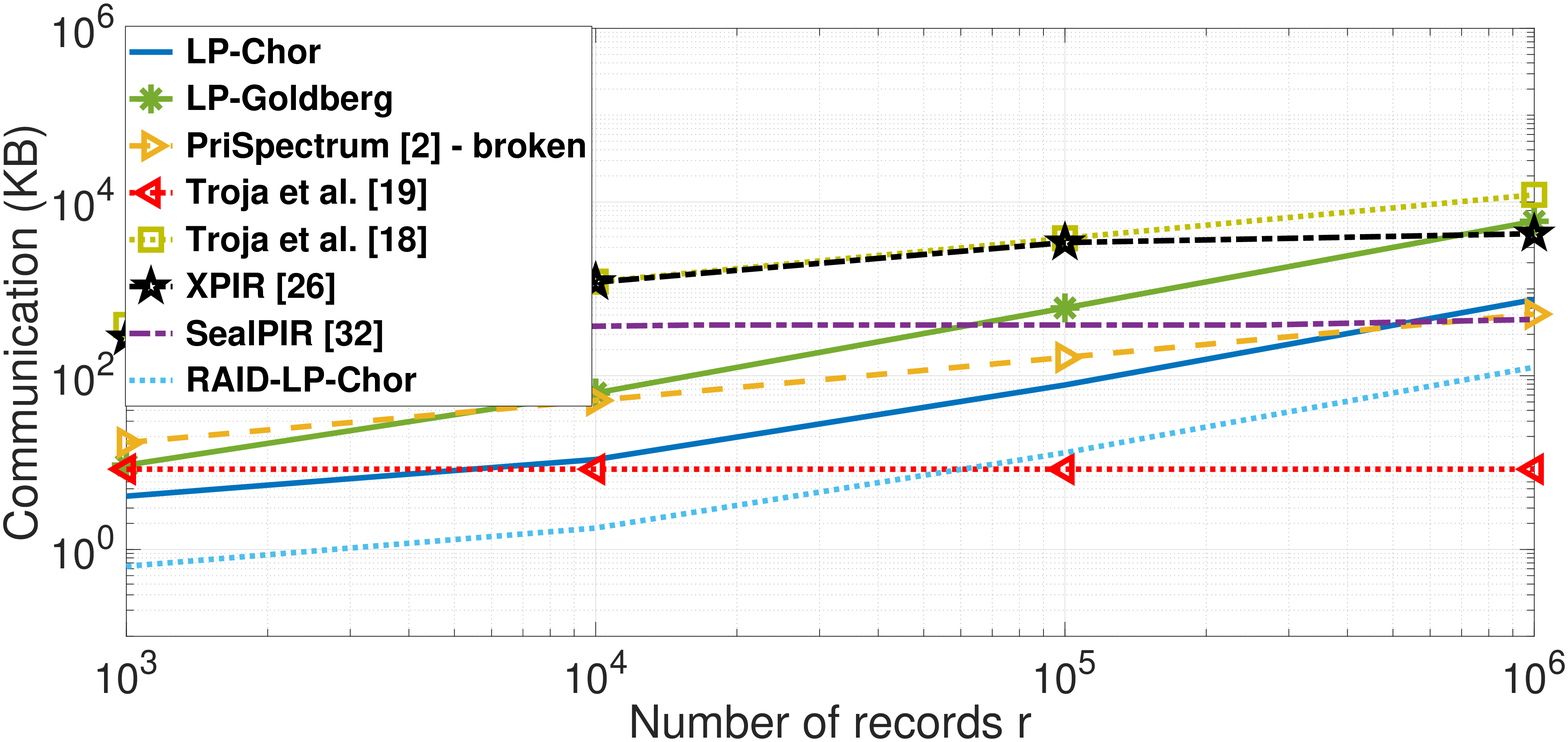}
 \caption{Comparison of the communication overhead of the different approaches: $\dbblock = 560$ B, $\kr = \ns$, $\vbr = 0$.}
\label{fig:comm}
 \vspace{-10pt}
\end{figure}

In terms of communication overhead, most of the approaches, including ours, have linear cost in the number of records in the database as shown in Table~\ref{tab:Table1}. What really makes a difference between these schemes' communication overheads is the associated constant factor which could be very large for some protocols. Based on our experiment and the expressions displayed in Table~\ref{tab:Table1}, we plot in \figurename~\ref{fig:comm}, the communication overhead that the \crn~experiences~for each private spectrum availability query issued by \su~for the different schemes. The scheme with the lowest communication overhead is that of Troja et al.~\cite{troja2015efficient} especially for a large number of records thanks to the use of Gentry et al. \pir~\cite{gentry2005single} which is the most communication efficient single-server protocol in the literature having a constant communication overhead. However this scheme is computationally expensive just like most of the existing \cpir-based approaches as we show in \figurename~\ref{fig:queryRTT}. $RAID\mhyphen$\chorScheme~is the second best scheme in terms of communication overhead followed by\chorScheme, but they also provide information theoretic privacy. As shown in Figure~\ref{fig:comm}, $RAID\mhyphen$\chorScheme~is significantly more efficient than \chorScheme, which again shows the benefit, in terms of overhead, of distributing the spectrum availability information among multiple \db s. As shown in \figurename~\ref{fig:comm}, \chorScheme~incurs much lower communication overhead than \goldbergScheme~thanks to the simplicity of the underlying Chor~\pir~protocol. However, as we discussed earlier, \goldbergScheme~provides additional security features compared to \chorScheme. SealPIR has a relatively high communication overhead especially for smaller database size but its overhead becomes comparable to that of \chorScheme~when the database's size gets larger as shown in \figurename~\ref{fig:comm}. This could be a good alternative to the \cpir~schemes used in the context of \crn s especially that it introduces much lower latency which is critical in the context of \crn s. Still, the proposed approaches have better performance and also provide information-theoretic privacy to \su s, which shows their practicality in real world. 

\section{Related Work}
\label{sec:Related}
There are other approaches that address the location privacy issue in database-driven \crn s. However, for the below mentioned reasons we decided not to consider them in our performance analysis.  For instance, Zhang et al.~\cite{zhang2015optimal} rely on the concept of {\em k-anonymity} to make each \su~queries \db~by sending a square cloak region that includes its actual location. {\em k-anonymity} guarantees that \su's location is indistinguishable among a set of $k$ points. This could be achieved through the use of dummy locations by generating $k-1$ properly selected dummy points, and performing $k$ queries to \db, using the real and dummy locations. Their approach relies on a tradeoff between providing high location privacy level and maximizing some utility. This makes it suffer from the fact that achieving a high location privacy level results in a decrease in spectrum utility. However, {\em k-anonymity}-based approaches cannot achieve high location privacy without incurring substantial communication/computation overhead. Furthermore, it has been shown in a recent study led by Sprint and Technicolor~\cite{zang2011anonymization} that anonymization based techniques are not efficient in providing location privacy guarantees, and may even leak some location information. Grissa et al~\cite{grissa2015cuckoo,grissa2017locationPrivacy} propose an information theoretic approach which could be considered as a variant of the trivial \pir~solution. They achieve this by using set-membership probabilistic data structures/filters to compress the content of the database and send it to \su~which then needs to try several combinations of channels and transmission parameters to check their existence in the data structure. However, LPDB is only suitable for situations where the structure of the database is known to \su s which is not always realistic. Also, LPDB relies on probabilistic data structures which makes it prone to false positives that can lead to erroneous spectrum availability decision and cause interference to \pu's transmission. 
Zhang et al.~\cite{zhang2015achieving} rely on the {\em $\epsilon$-geo-indistinguishability} mechanism~\cite{andres2013geo}, derived from {\em differential privacy} to protect bilateral location privacy of both \pu s and \su s, which is different from what we try to achieve in this paper. This mechanism helps \su s obfuscate their location, however, it introduces noise to \su's location which may impact the accuracy of the spectrum availability information retrieved.

\section{Conclusion}
\label{sec:Conclusion}
In this paper, with the key observation that database-driven \crn s contain multiple synchronized \db s having the same content, we harnessed multi-server \pir~techniques to achieve an optimal location privacy for both \su s and \pu s and for different use cases with high efficiency. Our analytical and experimental analysis indicates that our adaptation of multi-server \pir~for database-driven \crn s achieve magnitudes of time faster end-to-end delay compared to the fastest state-of-the-art single-server \pir~adaptation with an information theoretical privacy guarantee. Given the demonstrated benefits of multi-server \pir~approaches without incurring any extra architectural overhead on database-driven \crn s, we hope this work will provide an incentive for the research community to consider this direction when designing location privacy preservation protocols for \crn s. 
\section*{Acknowledgment}
This work was supported in part by the US National Science Foundation under NSF awards CNS-1162296 and CNS-1652389

\small{
\bibliographystyle{IEEEtran}
\bibliography{references,references-database,references-database-consistency,refs-security-privacy-sensing-14-15,refs-bechir-privacy-wireless-systems,references-TCCN-16-Mohamed}
}

\begin{IEEEbiography}
[{\includegraphics[width=1in,height=1.25in,clip,keepaspectratio]{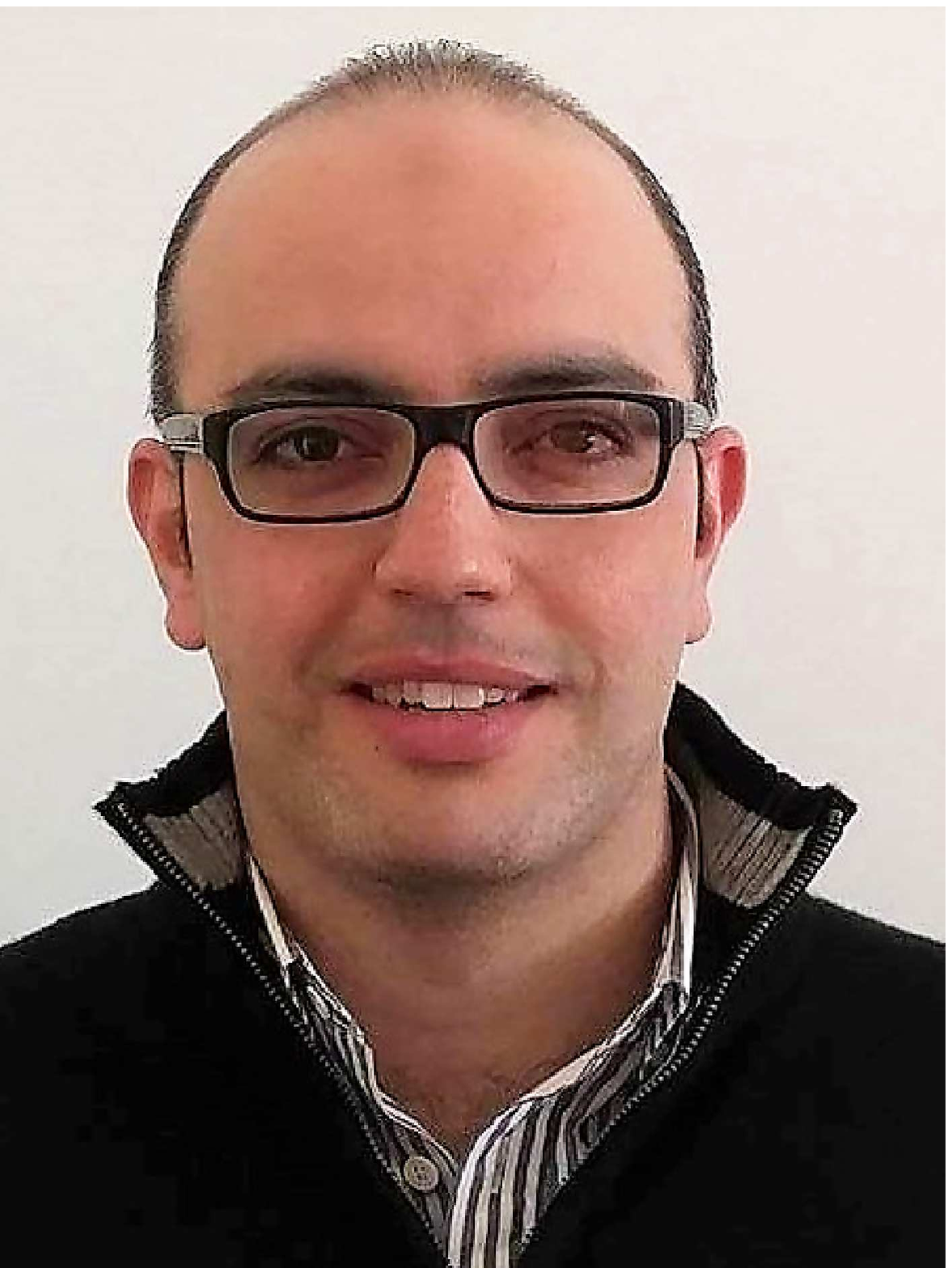}}]{Mohamed Grissa} (S'15) received the Diploma of Engineering (with highest distinction)
in telecommunication engineering from Ecole
Superieure des Communications de Tunis (Sup'Com), Tunis,
Tunisia, in 2011. He also received the
M.S. degree (June 2015) and the Ph.D. degree (September 2018) both in electrical and computer engineering (ECE) from Oregon State University, Corvallis, OR, USA.

Before joining Oregon State University, he worked as a Value Added Services Engineer at Orange France Telecom Group from 2012 to 2013. His research interests include privacy and security in computer networks, cognitive radio networks, spectrum access systems, IoT, Blockchain, and eHealth systems.
\end{IEEEbiography}

\begin{IEEEbiography}
[{\includegraphics[width=1in,height=1.25in,clip,keepaspectratio]{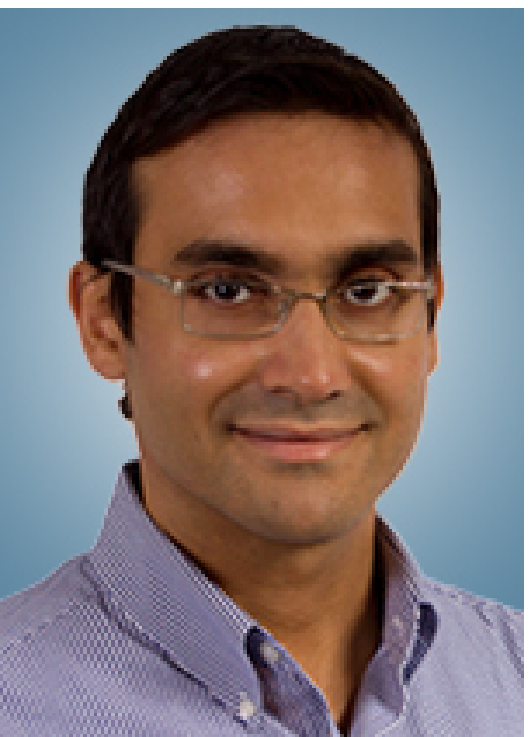}}]{Attila Altay Yavuz} (M'11) is an Assistant Professor in the Department of Computer Science and Engineering, University of South Florida  (2018). He was an Assistant Professor in the  School of Electrical Engineering and Computer Science, Oregon State University (2014-2018).
 He was a member of the security and privacy  research group at the Robert Bosch Research and Technology Center North America (2011- 2014). He received his PhD degree in Computer Science from North Carolina State University in August 2011. He received his MS degree in Computer Science from Bogazici University (2006) in Istanbul, Turkey. He is broadly interested in design, analysis and application of cryptographic tools and protocols to enhance the security of computer networks and systems. Attila Altay Yavuz is a recipient of NSF CAREER Award (2017). His research on privacy enhancing technologies (searchable encryption) and intra- vehicular network security are in the process of technology transfer with potential world-wide deployments. He has authored more than 40 research articles in top conferences and journals along with several patents. He is a member of IEEE and ACM.
\end{IEEEbiography}

\begin{IEEEbiography}
[{\includegraphics[width=1in,height=1.25in,clip,keepaspectratio]{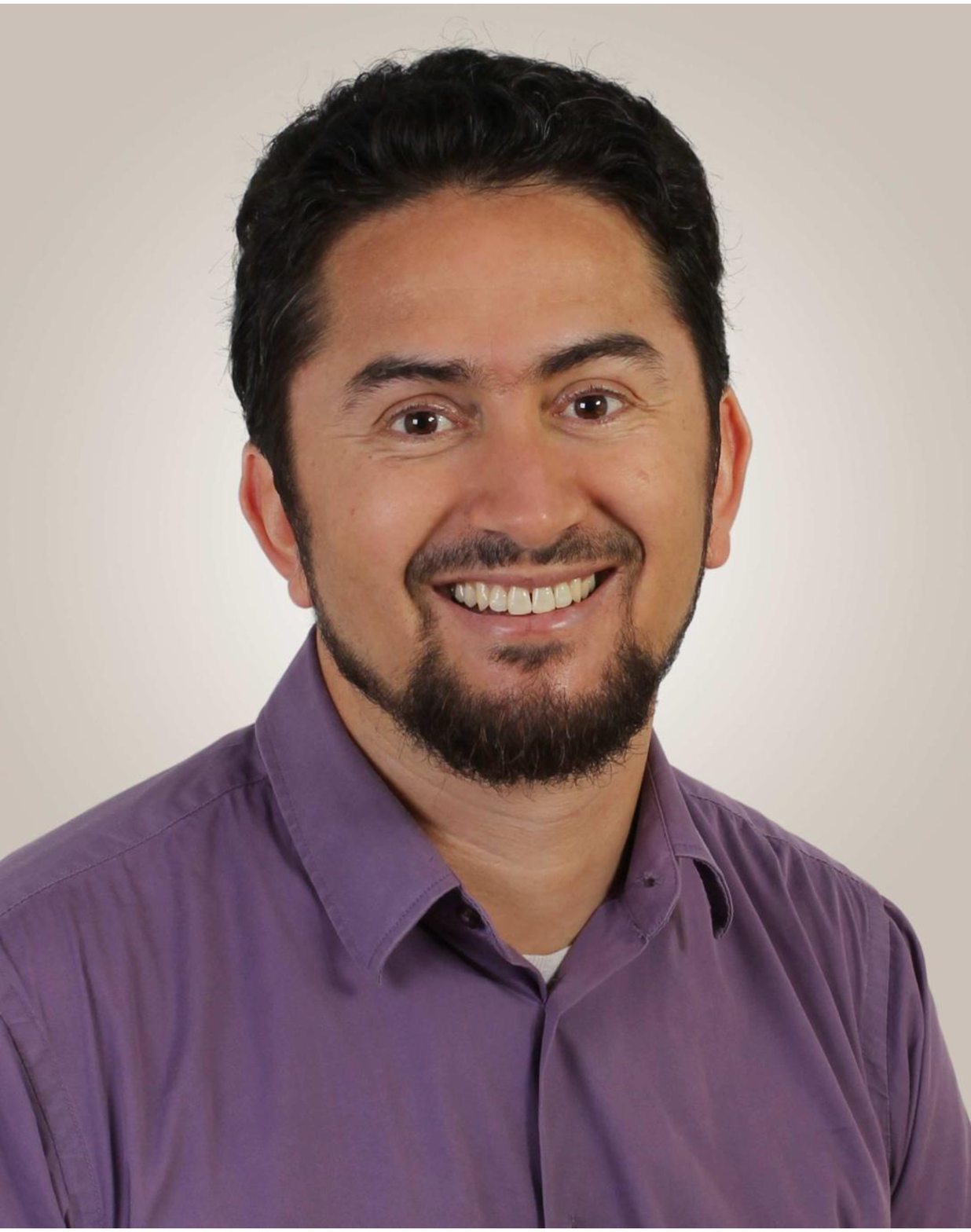}}]{Bechir Hamdaoui} (S'02\textendash M'05\textendash SM'12) is a Professor in the School of Electrical Engineering and Computer Science at Oregon State University. He received the Diploma of Graduate Engineer (1997) from the National School of Engineers at Tunis, Tunisia. He also received M.S. degrees in both ECE (2002) and CS (2004), and the Ph.D. degree in ECE (2005) all from the University of Wisconsin-Madison. His research interests are in the general fields of computer networking, mobile computing, and wireless communication, with a current focus on cloud computing, data analytics, distributed optimization and control, internet of things, cognitive radio and dynamic spectrum access, and security and privacy. He has won several awards, including the ICC 2017 Best Paper Award, the IWCMC 2017 Best Paper Award, the 2016 EECS Outstanding Research Award, and the 2009 NSF CAREER Award. He currently serves as Associate Editor for IEEE Transactions on Mobile Computing and for IEEE Network. He also served as Associate Editor for IEEE Transactions on Wireless Communications (2013-2018), IEEE Transactions on Vehicular Technology (2009-2014), Wireless Communications and Mobile Computing Journal (2009-2016), and Journal of Computer Systems, Networks, and Communications (2007-2009). He served as the chair for the 2017 INFOCOM Demo/Posters program, the 2016 IEEE GLOBECOM Mobile and Wireless Networks symposium, the 2014 IEEE ICC Communications Theory symposium, the 2011 ACM MOBICOM’s SRC program, and many other IEEE symposia and workshops, including ICC 2014, IWCMC 2009-2018, CTS 2012, and PERCOM 2009. He also served on technical program committees of many IEEE/ACM conferences, including INFOCOM, ICC, and GLOBECOM. He was selected and served as a Distinguished Lecturer for the IEEE Communication Society for 2016 and 2017. He is a Senior Member of IEEE, IEEE Computer Society, IEEE Communications Society, and IEEE Vehicular Technology Society.

\end{IEEEbiography}

\end{document}